\documentclass[twocolumn,preprint,10pt]{elsarticle}

\usepackage[utf8]{inputenc}
\usepackage[T1]{fontenc}
\usepackage{microtype}
\usepackage[htt]{hyphenat}

\usepackage{graphicx}
\usepackage{hyperref}
\usepackage{booktabs}
\usepackage{paralist}

\usepackage{xspace}

\usepackage[table]{xcolor}
\usepackage{multirow}
\usepackage{rotating}
\usepackage{listings} 
\usepackage{enumitem} 
\usepackage{ifthen}
\usepackage{amssymb}
\usepackage{flushend}

\usepackage{tabularx}
\usepackage{caption}

\usepackage{textcomp}

\usepackage{pgfplots}
\pgfplotsset{compat=newest}
 \usepackage{pgfplotstable}

\usepackage{adjustbox}
\usepackage{array}

\newcolumntype{R}[2]{%
    >{\adjustbox{angle=#1,lap=\width-(#2)}\bgroup}%
    l%
    <{\egroup}%
}
\newcommand*\rot{\multicolumn{1}{R{90}{1em}}}% no optional argument here, please!

\usepackage{todonotes}%[disable]

\newcommand{\squeeze}{\vspace{0\baselineskip}}%put between caption and graphics/table/whatever

\pgfplotstableread[col sep=comma]{
	smell, precision
	Subjective Language Smell, 0.96
	Ambiguous Adverbs and Adjectives Smell, 0.81
	Loophole Smell, 0.72
	Non-verifiable Term Smell, 0.70
	Superlative Requirements Smell, 0.49
	Comparative Requirements Smell, 0.48
	Negative Words Smell, 0.33
	Vague Pronouns Smell, 0.26
	Overall, 0.48
}\precisiondatatable

\makeatletter
\def\ps@pprintTitle{%
 \let\@oddhead\@empty
 \let\@evenhead\@empty
 \def\@oddfoot{Preprint accepted at Elsevier, \href{http://dx.doi.org/10.1016/j.jss.2016.02.047}{doi:10.1016/j.jss.2016.02.047}\hfill}%
 \let\@evenfoot\@oddfoot}
\makeatother

\begin{document}

\twocolumn[{
\begin{frontmatter}
  
\title{Rapid Quality Assurance with Requirements Smells}

\author[tum]{Henning Femmer\corref{cor1}}
\ead{femmer@in.tum.de}
\author[tum]{Daniel M\'{e}ndez Fern\'{a}ndez}
\ead{mendezfe@in.tum.de}
\author[ust]{Stefan Wagner}
\ead{Stefan.Wagner@informatik.uni-stuttgart.de}
\author[tum]{Sebastian Eder}
\ead{eders@in.tum.de}

\cortext[cor1]{Corresponding author}
\address[tum]{Software \& Systems Engineering, Technische Universit{\"a}t M{\"u}nchen, Germany} 
\address[ust]{Institute of Software Technology, University of Stuttgart, Germany}

%\maketitle
\begin{abstract} 
\textbf{Context:} Bad requirements quality can cause expensive consequences during the software development lifecycle, especially if iterations are long and feedback comes late. %-- the faster a problem is found, the cheaper it is to fix. This makes explicit the need of a lightweight detection mechanism of requirements quality violations. %, yet flexible enough to cope with the particularities of organisational contexts. 
\textbf{Objectives:} We aim at a light-weight static requirements analysis approach that allows for rapid checks immediately when requirements are written down. \textbf{Method:} We transfer the concept of \emph{code smells} to Requirements Engineering as \emph{Requirements Smells}. To evaluate the benefits and limitations, we define Requirements Smells, realize our concepts for a smell detection in a prototype called \emph{Smella} and apply Smella in a series of cases provided by three industrial and a university context. %We investigate whether and to which extent our approach supports the quality assurance of requirements artifacts. 
\textbf{Results:} 
%Our results strengthen our confidence in the be and our understanding of limitations of automatic Requirements Smell detection. 
The automatic detection yields an average precision of 59\% at an average recall of 82\% with high variation. The evaluation in practical environments indicates benefits such as an increase of the awareness of quality defects. Yet, some smells were not clearly distinguishable. \textbf{Conclusion:} Lightweight smell detection can uncover many practically relevant requirements defects in a reasonably precise way. Although some smells need to be defined more clearly, smell detection provides a helpful means to support quality assurance in Requirements Engineering, for instance, as a supplement to reviews.
\end{abstract}
\end{frontmatter}}]

\tableofcontents

% \todo[inline]{@Stefan: Explain reasonable precision}
%%%%%%%%%%%
%\section{todos}

%\todo[inline]{Consistency: in which order do we display smells? in all tables...} SE: Done. Subj., Amb., Loop., Non-Verif., Superl., Comp., Neg., Vag.
%\todo[inline]{Consistency: Smell names in texttt and examples in emph}
%\todo[inline]{Consistency: Smells vs Findings}
%%%%%%%%%%%%

%\todo[inline]{Check reference to figures and tables}
%\todo[inline]{Check bib references for completeness (year, pages, publisher etc}
\section{Introduction}

Defects in requirements, such as ambiguities or incomplete requirements, can lead to time and cost overruns in a pro\-ject~\cite{MW13b}.  %On the other hand, the notion of quality in general is multifaceted~\cite{1984_garvind_product_quality, KP96}, especially when it comes to the notion of requirements quality that depends on subjectivity and the context where the artifacts are used: A quality defect, such as ambiguity, might result in problems at one company while it might not be considered problematic at the next. 
%This highlights the importance of requirements quality assurance (QA) including reviews, which are time and cost intensive. 
Some of the issues require specific domain knowledge to be uncovered. For example, it is very difficult to decide whether a requirements artifact is complete without domain knowledge. Other issues, however, can be detected more easily: If a requirement states that a sensor should work with \emph{sufficient accuracy} without detailing what \emph{sufficient} means in that context, the requirement is vague and consequently not testable. The same holds for other pitfalls such as loopholes: Phrasing that a certain property of the software under development should be fulfilled \emph{as far as possible} leaves room for subjective (mis-)interpretation and, thus, can have severe consequences during the acceptance phase of a product~\cite{femmer2014rapid, ISO2011}.

To detect such quality defects, quality assurance processes often rely on reviews. Reviews of requirements artifacts, however, need to involve all relevant stakeholders~\cite{salger2013requirements}, who must manually read and understand each requirements artifact. Moreover, they are difficult to perform. They require a high domain knowledge and expertise from the reviewers~\cite{salger2013requirements} and the quality of their outcome depends on the quality of the reviewer~\cite{zelkowitz1983software}. On top of all this, reviewers could be distracted by superficial quality defects such as the aforementioned vague formulations or loopholes. We therefore argue that reviews are time-consuming and costly.
%Reviews, however, are error-prone~\cite{???} and cost-intensive~\cite{???}\todo{@all: references}, since often several review rounds are necessary to detect all quality defects. One reason for this could be the distraction of reviewers caused by superficial quality defects such as the afore-mentioned vague formulations or loopholes. 

 %%
%In addition, a problem that is found late in the project is, generally speaking, more expensive than if it was found early~\cite{Davis1993}. 
Therefore, quality assurance processes would benefit from faster feedback cycles in requirements engineering (RE), which support requirements engineers and project participants in immediately discovering certain types of pitfalls in requirements artifacts. Such feedback cycles could enable a lightweight quality assurance, e.g., as a complement to reviews. 

Since requirements in industry are nearly exclusively written in natural language~\cite{Luisa2004} and natural language has no formal semantics, quality defects in  requirements artifacts are hard to detect automatically. To face this challenge of fast feedback and the imperfect knowledge of a requirement's semantics, we created an approach that is based on what we call \emph{Requirements (Bad) Smells}. These are concrete symptoms for a requirement artifact's quality defect for which we enable rapid feedback through automatic smell detection.

In this paper, we contribute an analysis of whether and to what extent Requirements Smell analysis can support quality assurance in RE. To this end, we
\begin{compactenum}
\item define the notion of \emph{Requirements Smells} and integrate the Requirements Smells\footnote{In context of our studies, we use the ISO/IEC/\-IEEE 29148:2011 standard~\cite{ISO2011} (in the following: \emph{ISO~29148}) as basis for defining requirements quality. The standard supplies a list of so-called \emph{Requirements Language Criteria}, such as loopholes or ambiguous adverbs, which we use to define eight smells (see also the smell definition in Sect.~\ref{sec:smells_29148}).} concept into an analysis approach to complement (constructive and analytical) quality assurance in RE,
\item present a prototypical realization of our smell detection approach, which we call \emph{Smella}, and
\item conduct an empirical investigation of our approach to better understand the usefulness of a Requirements Smell analysis in quality assurance. 
\end{compactenum}

Our empirical evaluation involves three industrial contexts: The companies Daimler AG as a representative for the automotive sector, Wacker Chemie AG as a representative for the chemical sector, and TechDivison GmbH as an agile-specialized company. We complement the industrial contexts with an academic one, where we apply Smella to 51 requirements artifacts created by students. With our evaluations, we aim at discovering the accuracy of our smell analysis taking both a technical and a practical perspective that determines the context-specific relevance of the detected smells. We further analyze which requirements quality defects can be detected with smells, and we conclude with a discussion of how smell detection could help in the (industrial) quality assurance (QA) process. 

\subsection*{Previously published material}
This article extends our previously published workshop paper~\cite{femmer2014rapid} in the following aspects: We provide a richer discussion on the notion of Requirements Smell and give a precise definition. We introduce our (extended) tool-supported realization of our smell analysis approach and outline its integration into the QA process. We extend our first two case studies with another industrial one as well as with an investigation in an academic context to expand our initial empirical investigations by
\begin{compactenum}
\item investigating the accuracy of our smell detection including precision, recall, and relevance from a practical perspective,
\item analyzing which quality defects can be detected with smells and
\item gathering practitioner's feedback on how they would integrate smell detection in their QA process considering both formal and agile process environments.
\end{compactenum}

\subsection*{Outline}
The remainder of this paper is structured as follows. In Sect.~\ref{sec:rw}, we describe previous work in the area. In Sect.~\ref{sec:requirementsSmells}, we define the concept of Requirements Smells and describe how we derived a set of Requirements Smells from ISO~29148. We introduce the tool realization in Sect.~\ref{sec:smellDetection} and discuss the integration of smell detection in context of quality assurance in Sect.~\ref{sec:SmellsinQA}. In Sect.~\ref{sec:Evaluation}, we report on the empirical study that we set up to evaluate our approach, before concluding our paper in Sect.~\ref{sec:conclusion}.

%%%%%%%%%%%%%%%%%%%%%%%%%%%%%%%%%%%%%%%%%%%%%%%%%%%%%%%%%%%%%%%%%%%%%%%

\section{Related work}
\label{sec:rw}
In the following, we discuss work relating to the concept of natural language processing and smells in general, followed by quality assurance in RE, before critically discussing currently open research gaps.
%
%\subsection{Natural language processing}
%
%\todo[inline]{Various works on natural language processing, we refer the reader to Jurafsky and Martin, or others for the whole background.}
%\todo[inline]{To use in practice, various groups have worked on building comprehensive, usable libraries. E.g. Stanford Core NLP,  e.g. nltk for python, apache open nlp to name only a few}
%\todo[inline]{Some frameworks have been created to combine the various approaches: e.g. dkpro (building on UIMA), GATE, stanford NLP suite}
%\todo[inline]{For German, much less. For tokenization \cite{}, for pos tagging we use \cite{}, for lemmatization{}}
%It was not in the focus of this work to identify the best suitable of these, therefore, further improvement in precision (see RQ~2.1) might be possible.
%\todo[inline]{Beyond the work of this paper: coreference resolution, Word-Sense-Disambiguation, Grammar checking or proof reading (e.g. language tools), shallow semantic parsing,}

\subsection{The notion of smells in software engineering}
The concept of code smells was, to the best of our knowledge, first proposed by Fowler and Beck~\cite{Fowler1999a} to answer the question at which point the quality of code is so low that it must be refactored. According to Fowler and Beck, the answer cannot be objectively measured, but we can look for certain concrete, visible symptoms, such as duplicated code~\cite{Fowler1999a} as an indicator for bad maintainability~\cite{Juergens2009}. This concept of smells, as well as the list that Fowler and Beck proposed, led to a large field of research. Zhang et al.~\cite{Zhang2011} provide an in-depth analysis of the state of the art in code smells. The metaphor of smells as concrete symptoms has since then been transferred to quality of other artifacts including (unit) test smells~\cite{Deursen2001} and smells for system tests in natural language~\cite{Hauptmann2013a}. Ciemniewska et al.~\cite{Ciemniewska2007}, further characterize different defects of use cases through the term use case smell. In our work, we extend the notion of smells to the broader context of requirements engineering and introduce a concrete definition for the term \emph{Requirements Smell}.

\subsection{Quality assurance of software requirements}
The concept of Requirements Smells is located in the context of RE quality assurance (QA), which is performed either manually or automatically. 

\paragraph{Manual QA} Various authors have worked on QA for software requirements by applying manual techniques. Some put their focus on the classification of quality into characteristics~\cite{Davis1993}, others develop comprehensive checklists \cite{Anda2002, Berry2006, AxelvanLambsweerde,Kamsties2000,Kamsties2001}. Regarding QA, some develop constructive QA approaches, such as creating new RE languages, e.g.~\cite{denger2003higher}, to prevent issues up front, others develop approaches to make analytic QA, such as reviews, more effective~\cite{Shull2000}. In a recent empirical study on analytical QA, Parachuri et al.~\cite{Parachuri2014} manually investigate the presence of defects in use cases. To sum it up, these works on manual QA provide analytical and constructive methods, as well as (varying) lists for defects. They strengthen our confidence that today's requirements artifacts are vulnerable to quality defects.
 %\todo[inline]{DM: What where the results?} \todo[inline]{SW: I agree with DM. There needs to be some summarising statement what the state of manual QA is.}

%\todo{Comparatives and superlatives mentioned by \cite{Rupp2002}, manual checklist by Kamsties2000}

\paragraph{Automatic QA} Various publications discuss the automatic detection of quality violations in RE.
We summarize existing approaches and tools, their publications, and empirical evaluations in Table~\ref{tbl:related_work_eval}. We also created an in-depth analysis of in total 27 related publications evaluating which quality defects or smells the approaches opt for in their described detection. In the following, we will first explain two related areas (automatic QA for redundancy and for controlled languages), before discussing automatic QA for ambiguity in general. For ambiguity, we first describe those approaches that conducted empirical evaluations of precision or recall of quality defects related, but not identical to, the ones of ISO~29148.
Afterwards, we focus on publications that mention the same criteria as in the ISO~29148 (see Table~\ref{tbl:identical_smells} for this list and their respective empirical evaluations) and discuss the chosen approaches and results. We publish the complete list of each quality defect that is detected by each of the 27 papers, as well as the precision and recall (where provided), online as supplementary material~\cite{SmellaOnlineMaterial}.
%\todo[inline]{Use Trerm 'quality defect', and ISO~29148 criteria instead of 'our smells'}

\paragraph{Automatic QA for redundancy} One specific area of QA is avoiding redundancy and cloning. Whereas Juergens et al.~\cite{Juergens2010} use ConQAT to search for syntactic identity resulting from a copy-and-paste reuse, Falessi et al.~\cite{Falessi2013} aim at detecting similar content, therefore using methods from information retrieval (such as Latent Semantic Analysis~\cite{Lucia:2007:RTL:1276933.1276934}). Rago et al.~\cite{Rago2014} extend this work specifically for use cases. Their tool, ReqAlign, classifies each step with a semantic abstraction of the step. These publications analyze the performance of their approaches, and depending on the artifact and methods achieve precision and recall close to 1 (see Table \ref{tbl:related_work_eval}).

\paragraph{Automatic QA for controlled languages}  Another specific area is the application of controlled language and the QA of controlled language. RETA~\cite{Arora2015} specifically analyzes requirements that are written via certain requirements patterns (such as with the EARS template~\cite{Mavin2009}). Their goal is to detect both conformance to the template but also some of the ambiguities as defined by Berry et al~\cite{Berry2003}. The authors report on a case study where they look at the template conformance in depth, indicating that template conformance can be classified with various NLP suites to a high accuracy (Precision > 0.85, Recall > 0.9), both with and without glossaries. However, the performance of ambiguity detection (such as the detection of pronouns) is not further discussed in the publication. 
Similarly, AQUSA~\cite{Lucassen2015} analyzes requirements written in user story format (c.f.~\cite{cohn2004user} for a detailed introduction into user stories), and detects various defects, such as missing rationales, where they achieve a precision of 0.63-1.
Circe~\cite{Ambriola2006a, Gervasi2002} is a further tool that assumes that requirements are written in such requirements patterns  and detects violations of context- and domain-specific quality characteristics by building logical models. The authors report on six exemplary findings, which were detected in a NASA case study. However, despite their value to automatic QA, such approaches require very specific requirements structure. 

\paragraph{Automatic QA for ambiguity in general} The remaining approaches listed in Table~\ref{tbl:related_work_eval} aim at detecting ambiguities in unconstraint natural language.
 Since the quality defects detected by the approaches by Ciemniewska et al.~\cite{Ciemniewska2007}, Kof~\cite{Kof2007}, HeRA by Knauss et al.~\cite{Knauss2007a,Knauss2009b}, Kiyavitskaya et al.~\cite{Kiyavitskaya2008},  RESI by K\"orner et al.~\cite{Korner2009,Korner2009a,Korner2009b}, and Alpino by DeBruijn et al.~\cite{DeBruijn2010} are not the ones discussed in ISO~29148 and since we could not find an evaluation of precision and recall of these approaches, we omit discussing these approaches in-depth here. An analysis of what these approaches focus on in detail as well as their evaluation can be found in short in Table~\ref{tbl:related_work_eval} and in full length in our supplementary material online~\cite{SmellaOnlineMaterial}. In the following, we first report on those publications that focus on criteria different from ISO~29148, but which report precision or recall. Afterwards, we describe publications that aim at detecting quality violations of ISO~29148 (see Table~\ref{tbl:identical_smells}). 

First, Chantree et al.~\cite{Chantree2006} target the specific grammatical issue of coordination ambiguity (detecting problems of ambiguous references between parts of a sentence), mostly through statistical methods, such as occurrence and co-occurrence of words. In a case study, they report on a precision of their approach mostly between 54\% and 75\%. even though they do not explicitly differentiate between the detected ambiguities and the concept of pronouns. Second, Gleich et al.~\cite{Gleich2010} base their approach on the ambiguity handbook, as defined by Berry et al.~\cite{Berry2003}, as well as company-specific guidelines. They compare their dictionary- and POS-based approach against a gold standard which they created by letting people highlight ambiguities in requirements sentences. The gold standard deviates substantially, however, from what is considered high quality in their guidelines. Therefore, they create an additional gold standard, mostly based on the guideline rules. Consequently, their precision\footnote{Gleich et al. calculate their metrics based on the combination of all ambiguities; unfortunately, they do not differentiate, e.g. by the type of ambiguity. Also, to our knowledge, the gold standard does not differentiate between the types. This prevents a direct comparison to their work.} varies between 34\% for the pure experts opinion, to 97\% for a more guideline-based gold standard.
Third, Krisch and Houdek~\cite{Krisch2015}, focus on the detection of passive voice and so-called weak words. They present their dictionary- and POS-based approach to practitioners and find many false positives, similar to our RQ~3. In average, a precision of 12\% is reported for the weak words detection.
These approaches focus on very related, but not identical quality violations or smells.

\paragraph{Automatic QA for ISO~29148 criteria} Lastly, we specifically focus on those approaches that report to detect the criteria from the ISO~29148 standard. Table~\ref{tbl:identical_smells} provides an overview of these works and their respective evaluations.

The ARM tool~\cite{Wilson1997} defines quality in terms of the (now superseeded) IEEE~830 standard~\cite{IEEEComputerSociety1998} and proposes generic metrics, instead of giving feedback directly to requirements engineers. The metrics are calculated through counting how often a set of pre-defined terms (per metric) occurs in a document, including a metric of what we call Loopholes. Even though they report on a case study with 46 specifications from NASA, only a quantitative overview is reported\footnote{See also our RQ~1 in Sect.~\ref{sec:Evaluation}.}.
The QuARS tool~\cite{Fabbrini2001a,Fabbrini2001} is based on the author's experience. Bucchiarone et al.~\cite{Bucchiarone2005} describe the use of QuARS in a case study with Siemens and show some exemplary findings. SyTwo~\cite{Fantechi2003} adopts the quality model of QuARS and applies it to use cases. Loopholes and Subjectivity are part of the QuARS quality model.
Also RQA is built on a different, proprietary quality model, as described by G\'{e}nova et al.~\cite{Genova2011}, which includes negative terms as well as pronouns as quality defects.
 % They do not report on detailed results from case studies. %So far, an investigation of the application of the tool in practical contexts was not in scope of their research. 
These works also built upon extending natural language with NLP annotations, such as POS tags and searching through dictionaries for certain problematic phrases. However, we could not find a detailed empirical investigation of these tools, e.g.\ with regards to precision and recall.
SREE is an approach by Tjong and Berry~\cite{Tjong2013}, which aims at detection of ambiguities with a recall of 100\%. Therefore, they completely avoid all NLP approaches (since they come with imprecision), and build large dictionaries of words. The tool includes detection of loopholes, as well as pronouns; however, they report only on an aggregated precision for all the different types of ambiguities (66-68\%) from two case studies. 
In our previous paper~\cite{femmer2014rapid}, we searched for violations of ISO~29148, yet we provided only a quantitative analysis, as well as qualitative examples.
As mentioned before, RETA also issues warnings for pronouns, however, the evaluation in their paper~\cite{Arora2015} focusses on template conformance.
\begin{table*}[htbp]
\centering
\caption{Related work on criteria of ISO-29148 standard, detailed supplementary material can be found online~\cite{SmellaOnlineMaterial}}
\scriptsize
\label{tbl:identical_smells}
\begin{tabular}{@{}lccccccccc@{}}
\toprule
	&  ARM 	& \multicolumn{4}{c}{QuARS} 	& RQA 	&   SREE 	& Smella 	& RETA \\ 
 	%&  Wilson et al. 	& Fabbrini  et al. & Fabbrini et al.	& Fantechi  et al. 	& Bucchiarone et al. 	&  Genova2011 et al.	&   Tjong et al. 	& Femmer et al. 	& Arora et al. \\ 
	 	&  \cite{Wilson1997} 	& \cite{Fabbrini2001a} 	& \cite{Fabbrini2001} 	& \cite{Fantechi2003} 	& \cite{Bucchiarone2005} 	&  \cite{Genova2011} 	&  \cite{Tjong2013} 	& \cite{femmer2014rapid} 	& \cite{Arora2015} \\ 
\midrule
% 	& Quantiative analysis in number of findings is provided 	& Quantitative analysis and qualitative examples 	& Quantitative analysis and qualitative examples 	& Qualitative examples are provided 	& Qualitative examples are provided 	& Only solution proposals 	& Evaluation of precision 	& Qualitative examples are provided 	& Evaluation of precision and recall \\
Ambiguous Adv. \& Adj. 	&  	&  	&  	&  	&  	&  	&  	& E/Q 	&  \\
Comparatives 			&  	&  	&  	&  	&  	&  	&  	& E/Q 	&  \\
Loopholes (or Options) 	& Q 	&E/Q&E/Q& E 	& E 	&  	& Q*/P* 	& E/Q 	&  \\
Negative Terms 		&  	&  	&  	&  	&  	& O 	&  	& E/Q 	&  \\ 
Non-Verifiable Terms 	&  	&  	&  	&  	&  	&  	&  	& E/Q 	&  \\
Pronouns 				&  	& 	&  	&  	&  	& O 	& Q*/P* 	& E/Q 	& O \\
Subjectivity 			&  	&E/Q&E/Q& E 	& E 	&  	&  	& E/Q 	&  \\
Superlatives 			&  	&  	&  	&  	&  	&  	&  	& E/Q 	&  \\\bottomrule
\end{tabular}
\caption*{Legend: O=No empirical analysis, E=Examples from Case, Q=Quantification, P=Precision analyzed, R=Recall analyzed, *=Aggregated over multiple smells}
\end{table*}

\begin{sidewaystable*}
\centering
\caption{Related approaches and tools, and their evaluation, detailed supplementary material can be found online~\cite{SmellaOnlineMaterial}}%\scriptsize
\label{tbl:related_work_eval}
\begin{tabular}{@{}llllll@{}}
\toprule
Tool/Approach & Purpose (unless ambiguity det.) & Publications & Evaluation & Precision & Recall \\ \midrule
ConQAT & Redundancy & \cite{Juergens2010} & E/Q/P & 0.27-1 & -- \\
(Falessi) & Redundancy & \cite{Falessi2013} & Q/P/R & up to 96 & up to 96 \\
ReqAlign & Redundancy & \cite{Rago2014} & Q/P/R & 0.63 & 0.86 \\\midrule
RETA & Structured Language Rules & \cite{Arora2015} & E/Q/P/R & 0.85-0.94 & 0.91-1 \\
AQUSA & User Story Rules & \cite{Lucassen2015} & E/Q/P & 0.63-1 & -- \\
CIRCE & Structured Language Rules & \cite{Gervasi2002}  \cite{Ambriola2006a} & E & -- & -- \\\midrule
(Ciemniewska) &  & \cite{Ciemniewska2007} & E & -- & -- \\
(Kof) &  & \cite{Kof2007a} & E/Q & -- & -- \\
(Kiyavitskaya) &  & \cite{Kiyavitskaya2008} & E/Q & -- & -- \\
RESI &  & \cite{Korner2009} \cite{Korner2009a} \cite{Korner2009b} & E/Q & -- & -- \\
HeRA &  & \cite{Knauss2007a} \cite{Knauss2009b} & E & -- & -- \\
Alpino &  & \cite{DeBruijn2010} & E/Q & -- & -- \\\midrule
(Chantree) &  & \cite{Chantree2006} & E/P/R & 0.6-1 & 0.02-0.58 \\
Gleich &  & \cite{Gleich2010} & E/Q*/P*/R* & 0.34-0.97 & 0.53-0.86 \\
(Krisch) &  & \cite{Krisch2015} & E/Q/P & 0.12 & -- \\\midrule
ARM & RE Artifact Metrics & \cite{Wilson1997} & Q & -- & -- \\
QuARS / SyTwo &  & \cite{Fabbrini2001a} \cite{Fabbrini2001}  \cite{Fantechi2003} \cite{Bucchiarone2005} & E/Q & -- & -- \\
RQA &  & \cite{Genova2011} & O & -- & -- \\
SREE &  & \cite{Tjong2013} & Q*/P* & 0.66-0.68* & -- \\
Smella &  & \cite{femmer2014rapid} & E/Q & -- & -- \\ 
\bottomrule
\end{tabular}
\caption*{Legend: O=No empirical analysis, E=Examples from Case, Q=Quantification, P=Precision analyzed, R=Recall analyzed, *=Aggregated over multiple smells}
\end{sidewaystable*}

\subsection{Discussion}
Previous work has led to many valuable contributions to our field. To explore open research gaps, we now critically reflect on previous contributions from an evaluation, a quality definition and a technical perspective.

First, one gap in existing automatic QA ap\-proach\-es is the lack of empirical evidence, especially under realistic conditions. Only few of the introduced contributions were evaluated using industrial requirements artifacts. Those who do apply their approach on such artifacts focus on quantitative summaries explaining which finding was detected and how often it was detected. Some authors also give examples of findings, but only few works analyze this aspect in depth with precision and recall, especially in the fuzzy domain of ambiguity (see Table~\ref{tbl:related_work_eval}). When looking at the characteristics that are described in ISO~29148, we have not seen a quantitative analysis of precision and recall.
Furthermore, reported evidence does not include qualitative feedback from engineers who are supposed to use the approach, which could reveal many insights that cannot be captured by numbers alone. However, we postulate that the accuracy of quality violations very much depends on the respective context. This is especially true for the fuzzy domain of natural language where it is important to understand the (context-specific) impact of a finding to rate its detection for appropriateness and eventually justify resolving the issue. 
%A preliminary literature review comes to the same conclusion.\cite{Bano2015a}.

Second, the existing approaches are based on proprietary definitions of quality, based on experience or, sometimes, simply on what can be directly measured. The ARM tool~\cite{Wilson1997} is loosely based on the IEEE~830~\cite{IEEEComputerSociety1998} standard. However, as the recent literature survey by Schneider and Berenbach~\cite{Schneider2013} states: 
\emph{``the ISO/IEC/\-IEEE 29148:2011 is actually the standard that every requirements engineer should be familiar with''}.
We are not aware of an approach that evaluates the current ISO~29148 standard~\cite{ISO2011} in this respect. As Table~\ref{tbl:identical_smells} shows, for most language quality defects of ISO~29148, there has not yet been a tool to detect these quality defects. To all our knowledge, for neither of these factors, there is an differentiated empirical analysis of precision and recall.
%most factors of this standard have not been analyzed at all, let alone empiricall. %or, critically speaking, based on an explicitly defined understanding of how findings relate to quality or quality defects. Our approach of smells explicitly covers this aspect. 
Yet, many other quality models (most notably from the ambiguity handbook by Berry et al.~\cite{Berry2003}) and quality violations could lead to Requirements Smells, as far as they comply with the definition given in the next section.

Finally, taking a more technical perspective, our Requirements Smell detection approach does not fundamentally differ from existing approaches. Similar to previous works, we apply existing NLP techniques, such as lemmatization and POS tagging, as well as dictionaries. For the rules of the ISO~29148 standard, no parsing or ontologies (as used in other approaches) were required. However, to detect superlatives and comparatives in German, we added a morphological analysis, which have not yet seen in related work. 
%\todo[inline]{Add table }

In summary, in our contribution, we extend the current state of reported evidence on automatic QA for requirements artifacts via systematic studies in terms of distribution, precision, recall, and relevance, as well as by means of a systematic evaluation with practitioners under realistic conditions. We perform this on both existing, as well as new quality defects taken from the ISO~29148. Therefore, we extend our previously published first empirical steps~\cite{femmer2014rapid} to close these gaps by thorough empirical evaluation.

%\todo[inline]{And? Something is missing here.}
%
%\todo[inline]{Here should be a short description of the workshop paper and what we add in this paper.}
%in contrast to the previous study where we only looked at relevance \todo{abgrenzen}.

%In summary, we developed a systematic way to analyze for violations of the ISO~29148 natural language criteria and focus in our evaluation on qualitative feedback on the approach.

\section{Requirements Smells}
\label{sec:requirementsSmells}
We first introduce the terminology on Requirements Smells as used in this paper. In a second step, we define those smells we derived from ISO~29148 and which we use in our studies, before describing the tool realization in the next section. %We finally explain how we automatically detect the smells, present our prototypical implementation \emph{Smella}, and conclude with a brief discussion on its potential application in quality assurance.  

\subsection{Requirements Smell terminology}

Code smells are supposed to be an imprecise indication for bad code quality~\cite{Fowler1999a}. We apply this concept of smells to requirements and define it as follows:
A Requirements Smell is an indicator of a quality violation, which may lead to a defect, with a concrete location and a concrete detection mechanism. In detail, we consider a smell as having the following characteristics: 
\begin{enumerate}
	\item A Requirements Smell is an \emph{indicator} for a quality violation of a requirements artifact. For this definition, we understand requirements quality in terms of quality-in-use, meaning that bad requirements artifact quality is defined by its (potential) negative effects on activities in the software lifecycle that rely on these requirements artifacts (see also~\cite{FMM15}).
	%Bad (req.art.) quality is a fact in the requirements artifact that has a negative impact on a SE activity.
	\item A Requirements Smell does \emph{not necessarily lead to a defect} and, thus, has to be judged by the context (supported e.g.\ by (counter-/)indications). Whether a Requirements Smell finding is or is not a problem in a certain context must be individually decided for that context and is subject to reviews and other follow-up quality assurance activities.
	\item A Requirements Smell has a \emph{concrete location} in an entity of the requirements artifact itself, e.g.\ a word or a sequence. Requirements Smells always provide a pointer to a certain location that QA must inspect. In this regard, it differs from general quality characteristics, e.g.\ completeness, that only provide abstract criteria.
	\item A Requirements Smell has a \emph{concrete detection mechanism}. Due to its concrete nature, Requirements Smells offer techniques for detection of the smells. These techniques can, of course, be more or less accurate. 
\end{enumerate}

\begin{figure}[htbp]
\begin{center}
\includegraphics[width=1\columnwidth]{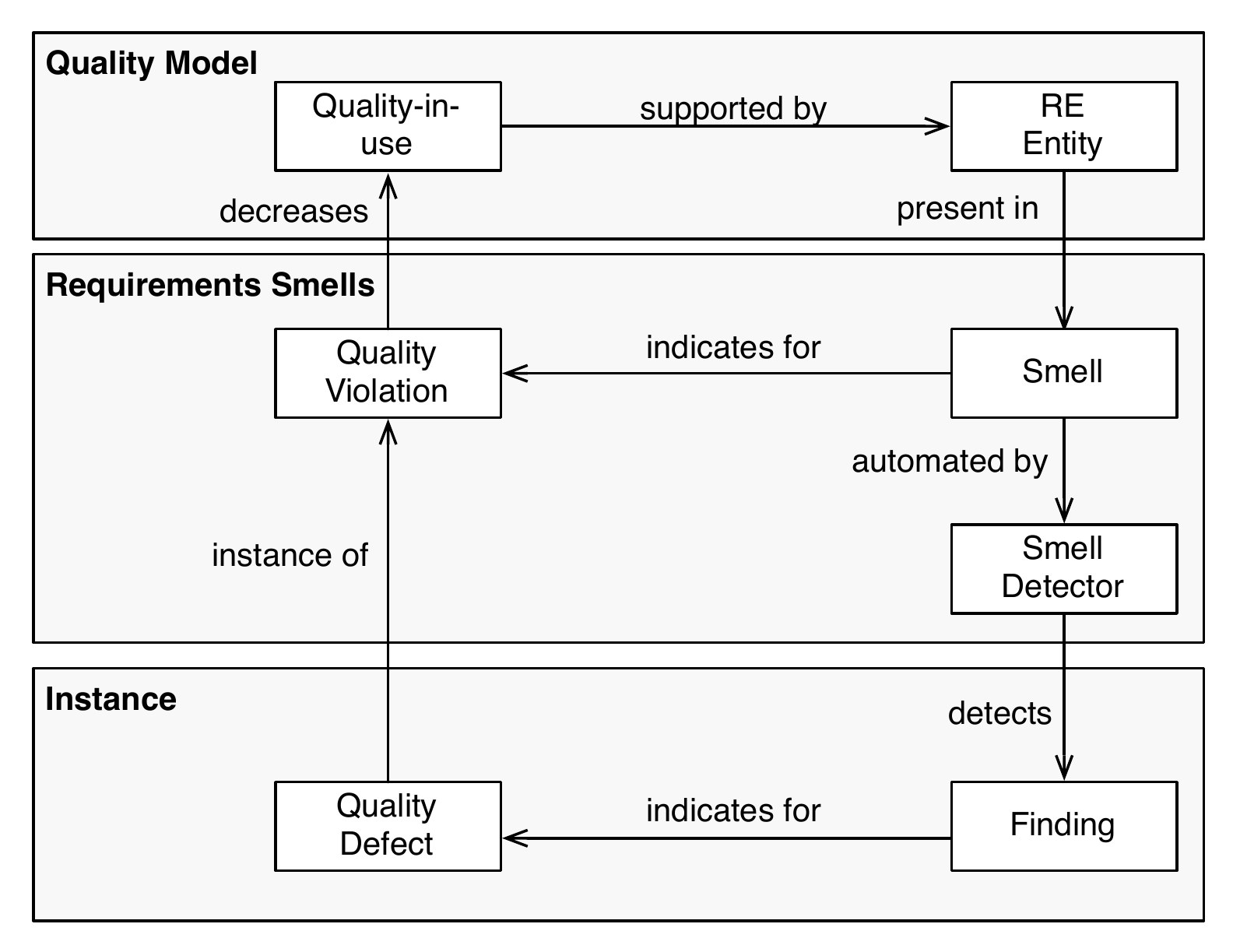}
\caption{Terminology of Requirements Smells (simplified)}
\label{fig:terminology}
\end{center}
\end{figure}

Furthermore, we define a \emph{quality defect} as a concrete instance or manifestation of a quality violation in the artifact, in contrast to a \emph{finding} which is an instance of a smell. However, like a smell indicates for a quality violation, the finding indicates for a defect. Fig.~\ref{fig:terminology} visualizes the relation of these terms.% as well as the given definition of Requirements Smells.

In the following, we will focus on natural language Requirements Smells, since requirements are mostly written in natural language~\cite{Luisa2004}. Furthermore, the real benefits of smell detection in practice should come with automation. Therefore, the remainder of the paper discusses only Requirements Smells where the detection mechanism can be executed automatically (i.e.\ it requires no manual creation of intermediate or supporting artifacts).

\subsection{Requirements Smells based on ISO~29148}
\label{sec:smells_29148}

We develop a set of Requirements Smells based on an existing definition of quality. For the investigations in scope of this paper, we take the ISO~29148 requirements engineering standard~\cite{ISO2011} as a baseline. The reasons for this are two-fold.
 
First, the ISO~29148 standard was created to harmonize a set of existing standards, including the IEEE~830:1998~\cite{IEEEComputerSociety1998} standard. It differentiates between quality characteristics for a set of requirements, such as completeness or consistency, and quality characteristics for individual requirements, such as unambiguity and singularity. The standard furthermore describes the usage of requirements in different project phases and gives exemplary contents and structure for requirements artifacts. Therefore, we argue that this standard is based on a broad agreement and acceptance. Recent literature studies come to the same conclusion~\cite{Schneider2013}. 

Second, the standard provides readers with a list of so-called \emph{requirements language criteria} which support the choice of proper language for requirements artifacts. The authors of the standard argue that violating the criteria results \emph{``in requirements that are often difficult or even impossible to verify or may allow for multiple interpretations"}~\cite[p.12]{ISO2011}. For defining our smells, which we describe next, we refer to this section of the standard and use all the defined requirements language criteria. We employ those criteria as a starting point and define the smells by adding the affected entities (e.g.\ a word) and an explanation. Here, we do not discuss the impact smells have on the quality-in-use. Essentially, smells hinder the understandability of requirements and consequently their subsequent handling and their verification (for a richer discussion, see also previous work in~\cite{FMM15}).

Our current understanding is based on the examples given by the standard. A subset of the language criteria, namely \texttt{Subjective Language}, \texttt{Ambiguous Adverbs and Adjectives} and \texttt{Non-verifiable Terms}, as defined in~\cite{ISO2011}, are strongly related, essentially since subjective language is a special type of ambiguity, which may lead to issues during verification. Since the intention of this work is to start with the standard as a definition of quality, in the following, we will remain with the provided definition based on the language criteria and leave the development of a precise and complete set of Requirements Smells to future work. In detail, we use the requirements language criteria to derive the smells summarized next.

\vspace{0.5cm}
\newcommand{\defWidth}{5cm}

\begin{tabular}{l p{\defWidth}}\toprule
Smell Name: & \texttt{Subjective Language} \\
Entity: & Word \\
Explanation: & Subjective Language refers to words of which the semantics is not objectively defined, such as \emph{user friendly}, \emph{easy to use}, \emph{cost effective}. \\ 
Example: & The architecture as well as the programming must ensure a \textbf{simple} and \textbf{efficient} maintainability.\\
%Quality Relation: & Each person using the artifact might develop his own interpretation of when this requirement is fulfilled, which will get visible especially during testing, when the requirements engineers', developer's and tester's understanding are matched against each other. Therefore, this smell has, among other impacts, a negative impact on understanding requirements, as well as defining test cases.\\
%Detection Mechanism: & Dictionary\\
\bottomrule
\end{tabular}

\vspace{0.5cm}

\begin{tabular}{l p{\defWidth}}\toprule
Smell Name: & \texttt{Ambiguous Adverbs and Adjectives} \\
Entity: & Adverb, Adjective \\
Explanation: & Ambiguous Adverbs and Adjectives refer to certain adverbs and adjectives that are unspecific by nature, such as \emph{almost always}, \emph{significant} and \emph{minimal}.\\ 
Example: & If the (...) quality is \textbf{too low}, a fault must be written to the error memory.\\
%Quality Relation: & Like above, this smell has a negative impact on understanding requirements, as well as defining test cases.\\
%Detection Mechanism: & Dictionary\\
\bottomrule
\end{tabular}

\vspace{0.5cm}

\begin{tabular}{l p{\defWidth}}\toprule
Smell Name: &  \texttt{Loopholes}\\
Entity: & Word \\
Explanation: & Loopholes refer to phrases that express that the following requirement must be fulfilled only to a certain, imprecisely defined extent. \\
%Examples: \emph{if possible}, \emph{as appropriate}, \emph{as applicable} \\ 
Example: & \textbf{As far as possible}, inputs are checked for plausibility.\\
%Quality Relation: & With Loopholes, stakeholders can (intentionally or unintentionally) ignore certain parts of the requirements artifact. In a context of legal binding of requirements, loopholes can create a gap between intention and implementation. This negatively impacts various activities including, inter alia, the validation of a solution.\\
%Detection Mechanism: & Dictionary \\ 
\bottomrule
\end{tabular}
%\todo[inline]{Ist der impact Vorschlag fuer Loopholes ok? (DM)}
%\todo[inline]{hab noch 2, satz hinzugefuegt (HF)}

\vspace{0.5cm}

\begin{tabular}{l p{\defWidth}}\toprule
Smell Name: & \texttt{Open-ended, Non-verifiable Terms} \\
Entity: & Word \\
Explanation: & Open-ended, non-verifiable terms are hard to verify as they offer a choice of possibilities, e.g. for the developers.\\
%Examples: \emph{provide support}, \emph{but not limited to}, \emph{as a minimum}\\ 
Example: & The system may only be activated, if all required sensors (...) work with \textbf{sufficient} measurement accuracy.\\
%Quality Relation: & Open-ended and non-verifiable terms can lead to requirements that are difficult to test or even impossible to determine whether or not they are fulfilled for a software. This negatively impacts defining test cases. \\
%Detection Mechanism: & Dictionary\\
\bottomrule
\end{tabular}

\vspace{0.5cm}

\begin{tabular}{l p{\defWidth}}\toprule
Smell Name: & \texttt{Superlatives} \\
Entity: & Adverb, Adjective \\
Explanation: & Superlatives refer to requirements that express a relation of the system to all other systems.\\
%Examples: \emph{best performance}, \emph{lowest response time}.
Example:& The system must provide the signal in the \textbf{highest} resolution that is desired by the signal customer.\\
%Quality Relation: & Defining a system in relation to others creates a relative definition of fulfillment, e.g. \emph{fastest on the market}. Since this creates a potentially moving target, this smell has a negative impact on defining and executing reliable test cases.\\
%Detection Mechanism: & Morphological analysis or POS tagging \\
\bottomrule
\end{tabular}

\vspace{0.5cm}

\begin{tabular}{l p{\defWidth}}\toprule
Smell Name: & \texttt{Comparatives} \\
Entity: & Adverb, Adjective\\
Explanation: & Comparatives are used in requirements that express a relation of the system to specific other systems or previous situations. \\ 
Example: & The display (...) contains the fields A, B, and C, as well as \textbf{more exact} build infos.\\
%Quality Relation: & Defining a system in relation to others creates a relative definition of fulfillment, e.g. \emph{faster than X}. Since this creates a potentially moving target, this smell has a negative impact on defining and executing reliable test cases.\\
%Detection Mechanism: & Morphological analysis or POS tagging \\
\bottomrule
\end{tabular}

\vspace{0.5cm}

\begin{tabular}{l p{\defWidth}}\toprule
Smell Name: & \texttt{Negative Statements} \\
Entity: & Word \\
Explanation: & Negative Statements are ``statements of system capability not to be provided"\cite{ISO2011}. Some argue that negative statements can lead to underspecification, such as lack of explaining the system's reaction on such a case.\\
%Example: \emph{The system must not accept VISA credit cards.}
%For this example, a more complete specification describes how the system reacts on the unaccepted input. \\ 
Example: & The system must \textbf{not} sign off users due to timeouts.\\
%Quality Relation: &  A system could be easier to understand, and especially implemented, if the artifact describes what the system should do instead of what it should not do. Hence, this smell might have a negative impact on understanding and, especially, implementing.\\
%Detection Mechanism: & POS tagging or dictionary\\
\bottomrule
\end{tabular}

\vspace{0.5cm}

\begin{tabular}{l p{\defWidth}}\toprule
Smell Name: & \texttt{Vague Pronouns} \\
Entity: & Pronoun \\
Explanation: & Vague Pronouns are unclear relations of a pronoun.\\
Example: & The software must implement services for applications, \textbf{which} must communicate with controller applications deployed on other controllers. \\
%
%[Note: The translation is less ambiguous than the original finding in German, as the reflexive pronoun in English identifies its relation more clearly. The original requirement stated: \emph{Die Software muss Dienste f\"{u}r Anwendungen implementieren, welche \"{u}ber ein Steuerger\"{a}t hinaus mit anderen Steuerger\"{a}te-Anwendungen kommunizieren m\"{u}ssen.}] 
%Quality Relation: & Vague pronouns make it harder to read and comprehend the requirements, therefore this smell might have a negative impact on understanding.\\
%Detection Mechanism: & POS tagging: Finding substituting pronouns.\\
\bottomrule
\end{tabular}

\vspace{0.5cm}

\begin{tabular}{l p{\defWidth}}\toprule
Smell Name: &  \texttt{Incomplete References}\\
Entity: & Text reference \\
Explanation: & Incomplete References are references that a reader cannot follow (e.g.\ no location provided).\\
Example:& \emph{[1] ``Unknown white paper". Peter Miller.}\\ 
%Quality Relation: &  With incomplete references, the reader might not be able to follow the artifact, therefore this smell might have a negative impact on understanding \\
%Detection Mechanism: & \\
\bottomrule
\end{tabular}
\vspace{0.5cm}

%We implemented all these smells but \emph{incomplete references} as there were no explicit references in our requirements artifacts. 
%
%\todo[inline]{DM: What exactly does this mean? That we realized those smells in the tool but incomplete refs because our cases in the studies had no references? I wouldn't formulate it in a way such that the tool would be only a speific (study) purpose development (also indicated by last sentence in 4.3)} 
% All remaining features were considered as smells for bad quality of requirements specifications. 
%At this point we assume that these smells apply for all specifications; however, we will discuss the appropriateness of the given list based on concrete experience from the case studies in Sect.~\ref{sec:discLangCrit}.\todo{Check later, if still part of discussion}

%\todo[inline]{@Henning: Discuss fuzziness, differentiation, etc.}

\section{Smella: A prototype for Requirements Smell detection}
\label{sec:smellDetection}
Requirements Smell detection, as presented in this paper, serves to support manual quality assurance tasks (see also the next section). The smell detection is implemented on top of the software quality analysis toolkit ConQAT,\footnote{\url{http://www.conqat.org}} a platform for source code analysis, which we extended with the required NLP features. In the following, we introduce the process for the automatic part of the approach, i.e.\ the detection and reporting of Requirements Smells. To the best of our knowledge, there is no tool, other than the ones mentioned in related work, that detect and present these smells in natural language requirements documents.

\begin{figure*}[htbp]
\begin{center}
\includegraphics[width=\textwidth]{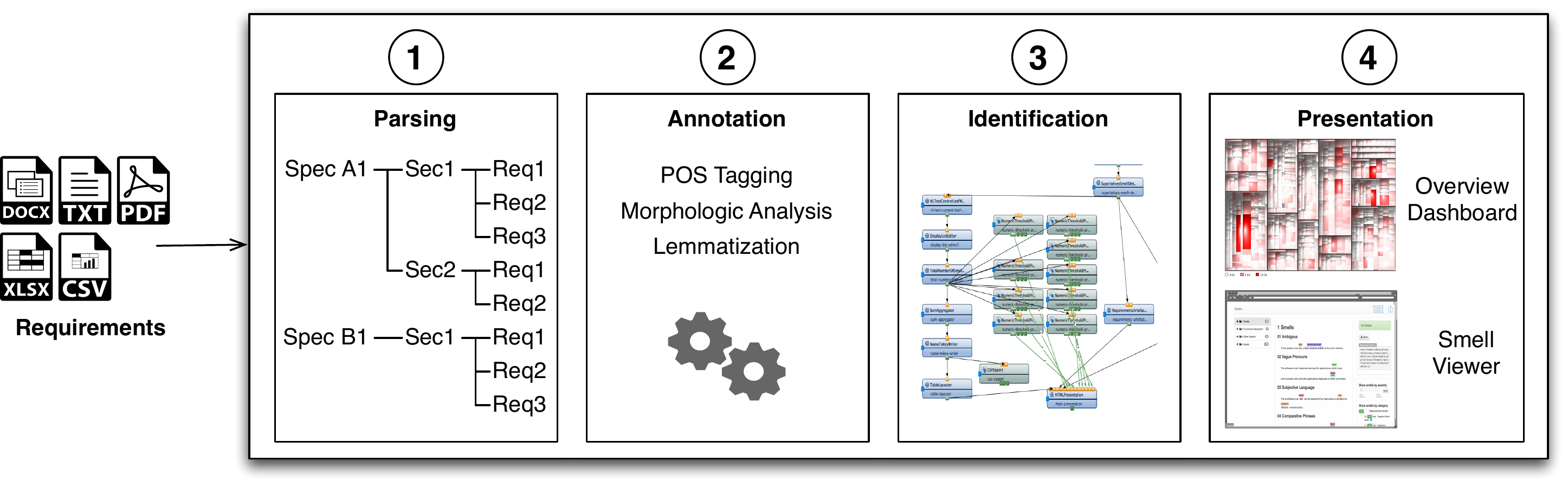}\squeeze
\caption{The overall smell detection process}
\label{fig:analysis}\squeeze
\end{center}
\end{figure*}

The process takes requirements artifacts in various formats (MS Word, MS Excel, PDF, plain text, comma-separated values) and consists of four steps (see also Fig.~\ref{fig:analysis}): 
\begin{enumerate}
  \item \emph{Requirements parsing} of the requirements artifacts into single items (e.g. sections or rows), resulting in plain texts, one for each item
  \item \emph{Language annotation} of the requirements with meta-information
  \item \emph{Findings identification} in the requirements, based on the language annotations% \todo{identification vs annotation} 
  \item \emph{Presentation} of a human-readable visualization of the findings as well as a summary of the results
\end{enumerate}
%Finally, the findings are displayed as an integrated part in the requirements artifact. 
The techniques behind these steps are explained in the following section.

\subsection{Requirements parsing}
%\todo[inline]{FILL}
%\todo[inline]{Done?}
Our current tool is able to process several file formats: MS Word, MS Excel, PDF, plain text and comma-separated values (CSV). Depending on the format, the files are parsed in different ways. Plain text and PDF are taken as is and parsed file by file. Microsoft Word files are grouped by their sections. For Microsoft Excel and CSV files, we define those columns that represent the IDs or names (if there are any), and those columns should be used as text input to detect smells. 

If a file is written in a known template, such as a common template for use cases, we can make use of this template to understand structural defects, such as lacking content items in a template. In the remainder of this paper, however, we focus on the natural language Requirements Smells as provided by the ISO standard. 

%Regarding the counting of artifacts, we relayed on the structure provided in the artifacts, since the differences between the various concepts and templates such as unstructured requirements, use cases, user stories etc. do not allow to translation from one size metric into the other. Therefore, we quantify the size of an artifact in terms of its 

%\begin{itemize}
%	\item [plain text:] We simply take the text (as is) from the file.
%	\item [MS Word:] We omit meta information as headers and footers, as well as the version history of the files. But wwe take the structure of these files into account, if it follows a given template. For example, when parsing artefacts containing use cases, we identify singe steps in the main and alternative flows, and detect references between these steps. Furthermore, we identify the different parts of the use case, e.g., the flows, the section containing stakeholders, or the rationale.
%	\item [MS Excel:] \todo[inline]{Was machen wir da?}
%	\item [comma-separated values] \todo[inline]{Und hier?}
%	\item [PDF] \todo[inline]{Hier weiss ich es auch nicht...?}
%\end{itemize}

\subsection{Language annotation}
\label{sec:nlp}

For the annotation and smell detection steps, we employ three techniques from Natural Language Processing (NLP) \cite{Jurafsky2014}. Table~\ref{tbl:smellImplementation} additionally shows which of the techniques we use for which smell.

\begin{description}
  \item[POS Tagging:] For two smells, we use part-of-speech (POS) tagging. Given a sentence in natural language, it determines the role and function of each single word in the sentence. The output is a so-called \emph{tag} for each word indicating, for instance, whether a word is an adjective, a particle, or a possessive pronoun. We used the Stanford NLP library \cite{Toutanova2003} and the RFTagger~\cite{Schmid2008} for this. Both are statistical, probabilistic taggers that train models similar to Hidden Markov Models based on existing databases of tagged texts. A detailed introduction into the technical details of POS tagging is beyond the scope of this paper but can be found, for example, in~\cite{Jurafsky2014}. We use POS tagging to determine so-called substituting pronouns. These are pronouns that do not repeat the original noun and, thus, need a human's interpretation of its dependency.
  \item[Morphological Analysis:] Based on POS tagging, we perform a more detailed analysis of text and determine a word's inflection. This includes, inter alia, determining a verb's tense or an adjective's comparison. We use this technique to analyze if adjectives or adverbs are used in their comparative or superlative form.
    \item[Dictionaries \& Lemmatization:] For the remaining five smells, we use dictionaries based on the proposals of the standard \cite{ISO2011} and on our experiences from first experiments in a previous work~\cite{femmer2014rapid}. We furthermore apply lemmatization for these words, which is a normalization technique that reproduces the original form of a word. In other words, if a lemmatizer is applied to the words \emph{were}, \emph{is} or \emph{are}, the lemmatizer will return for all three the word \emph{be}. Lemmatization is in its purpose very similar to stemming (see, e.g.\ the Porter Algorithm~\cite{Porter1980}), yet not based on heuristics but on the POS tag as well as the word's morphological form. For Requirements Smells, the difference is significant: For example, the words \emph{use} and \emph{useful} stem to the same word origin (\emph{use}), but to different lemmas (i.e. meanings;  \emph{use} and \emph{useful}). Whereas the lemma \emph{use} is mostly clear to all stakeholders, the lemma \emph{useful} is easily misinterpreted. 
\end{description}

\subsection{Findings identification}
Based on the aforementioned information, we identify findings. This step actually finds the parts of an artifact that exhibit bad smells. Dependent on the actual smell, we use different techniques, as shown in Table~\ref{tbl:smellImplementation}. If the smell relates to a grammatical aspect, we search through the information from POS tagging and morphological analyses. For example, for the Superlatives Smell, we report a finding if an adjective is, according to morphologic analysis, inflected in its superlative form. If the smell does not relate to grammatical aspects but rather the semantics of the requirements, we identify the smell by matching the lemma of a word against a set of words from pre-defined dictionaries.
Since the requirements under analysis in our cases did not contain references, incomplete references are not part of our tool at present.

\begin{table*}[ht!]
\caption{Detection techniques for smells}\label{tbl:smellImplementation}
\centering
\begin{tabular}{p{6cm}p{7cm}}\toprule
Smell Name                        & Detection Mechanism \\
\midrule
Subjective Language               & Dictionary\\
Ambiguous Adverbs and Adjectives  &	Dictionary\\
Loopholes                         & Dictionary\\
Open-ended, non-verifiable terms  & Dictionary\\
Superlatives                      & Morphological analysis or POS tagging \\
Comparatives                      & Morphological analysis or POS tagging \\
Negative Statements               & POS tagging and dictionary\\
Vague Pronouns                    & POS tagging: Substituting pronouns.\\
Incomplete References             & Not in scope of this study\\
\bottomrule
\end{tabular}
\end{table*}

%\todo[inline]{Remove tool-specific discussions from this section}

\begin{figure*}[htbp]
\begin{center}
\includegraphics[width=\textwidth]{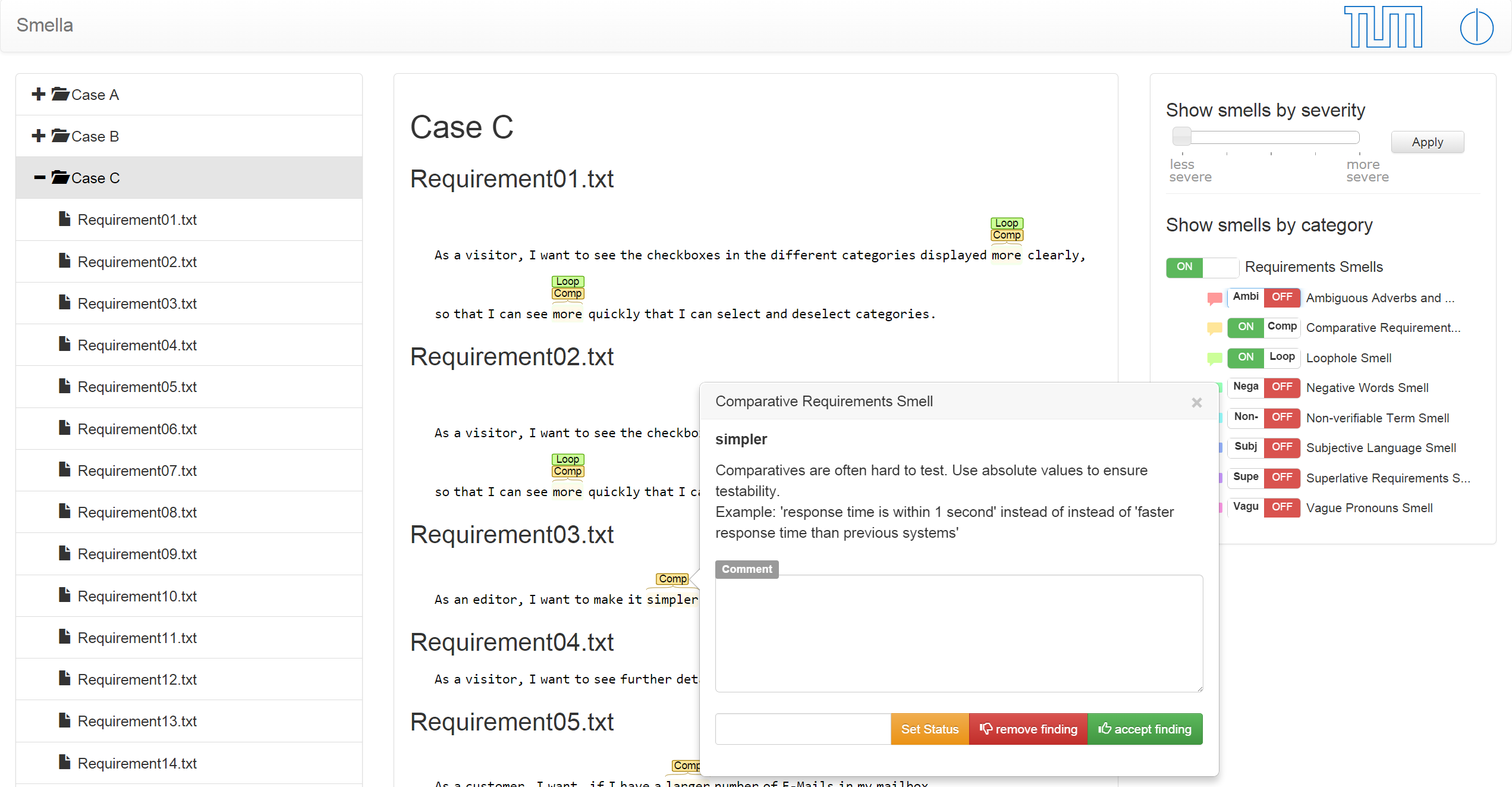}\squeeze
\caption{A sample output from the smell detection tool (detailed artifact view) with some smells disabled and some findings blacklisted}\squeeze
\label{fig:tool}
\end{center}
\end{figure*}

\begin{figure*}[htbp]
\begin{center}
\includegraphics[width=\textwidth]{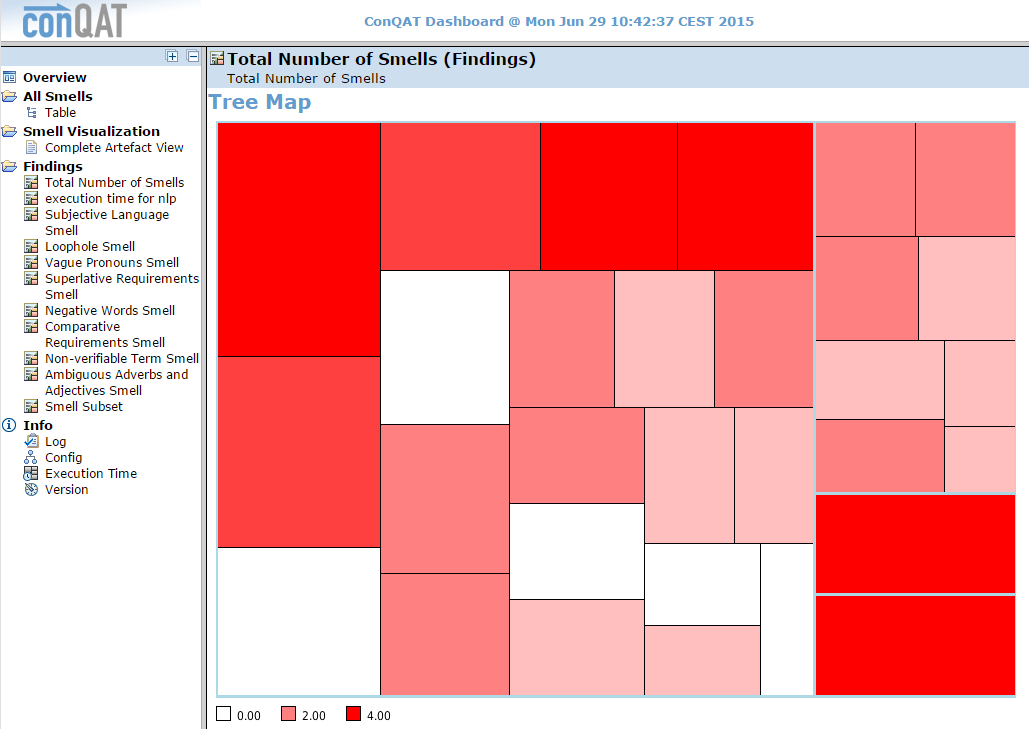}\squeeze
\caption{A sample output from the smell detection tool (hotspot analysis view)}\squeeze
\label{fig:conqat}
\end{center}
\end{figure*}

\subsection{Findings presentation}
\label{sec:Smella}

%\todo[inline]{Screenshots}
%\todo[inline]{maybe subsubsec core features, then description, dann subsubsec UI?}

We implemented the presentation of findings in a prototype, which we call \emph{Smella} (Smell Analysis). Smella is a web-based tool that enables viewing, reviewing and blacklisting findings as well as a hotspot analysis at an artifact level. In the Smella presentation, we display the complete requirements artifact and annotate findings in a spell checker style. This follows the idea of smells as only indications that must be evaluated in their context. Lastly, the system gives detailed information when a user hovers a finding (see Fig.~\ref{fig:tool}). In the following, we shortly describe the features of Smella in detail to provide the reader with a rough understanding of the prototype. 

\begin{description}
\item[View findings:] At the level of a single artifact, we present the text of the artifact and its structure. We mark all findings in the text. With a click on the markers, more information about the finding is displayed. The tool provides an explanation of the rationale behind this smell and possible improvements for the finding depending on the smell (every smell has a message for improvements).

\item[Review findings:] We allow the user to write a review and to set a status for each finding, both supporting feedback mechanisms within and between project teams. A user has the possibility to accept or reject a finding but also to set a custom state, for example \emph{under review}. Accepting a finding means the finding needs to be addressed. If a finding is rejected, the finding does not need to be addressed. The semantics of the custom status is open to the reviewer.

\item[Blacklist findings:] Smells are only indicators for issues. Therefore, users can reject findings. If a finding is rejected by the user, the finding is removed from the visualization and will not be presented to the user anymore.

\item[Disable smells:] Often, users are interested in only a subset of smells or even just one smell. Therefore, we allow the user to hide all findings of particular smells and to select the smells she wants to display in the artifact view.

\item[Analyze hotspots:] In this view, we present all artifacts in a colored treemap (see Fig.~\ref{fig:conqat}). Every box in the treemap is one artifact, with the color of the box indicating the number of findings: the more red an artifacts is, the more findings it contains (the more it ``smells'' bad). The artifacts are grouped by their folder structure. The tool provides a summarized treemap for all smells as well as a separate treemap for all individual smells. With these treemaps, users can identify artifacts or groups of artifacts exhibiting a high number of findings -- for one single smell but also for all smells together. This feature supports the identification of candidates for in-depth reviews.
\end{description}
%
%\todo[inline]{Describe how it can be used, what are the main features}
%\begin{itemize}
%\item feedback resulting from discussion with experts: what's wrong, what should be done in a better way
%\item showing only types of issues
%\item blacklisting/hiding
%\item hotspot view
%\item ?
%\end{itemize}

%\todo[inline]{Describe how the UI is structured etc.}
%\todo[inline]{Integrate/separate previous section}

\section{Requirements Smell detection in the process of quality assurance}
\label{sec:SmellsinQA}

The Requirements Smell detection approach described in previous sections serves the primary purpose of supporting quality assurance in RE. The detection process itself is, however, not restricted to particular quality assurance tasks, nor does it depend on a particular (software) process model as we will show in Sect.~\ref{sec:Evaluation}. Hence, a smell detection, similar to the notion of quality itself, always depends on the views in a socio-economic context. Thus, how to integrate smell detection into quality assurance needs to be answered according to the particularities of that context. In the following, we therefore briefly outline the role smell detection can generally take in the process of quality assurance. More concrete proposals on how to integrate it into specific contexts are given in our case studies in Sect.~\ref{sec:Evaluation}.

\begin{figure*}[hbt]
\begin{center}
\includegraphics[width=.8\textwidth]{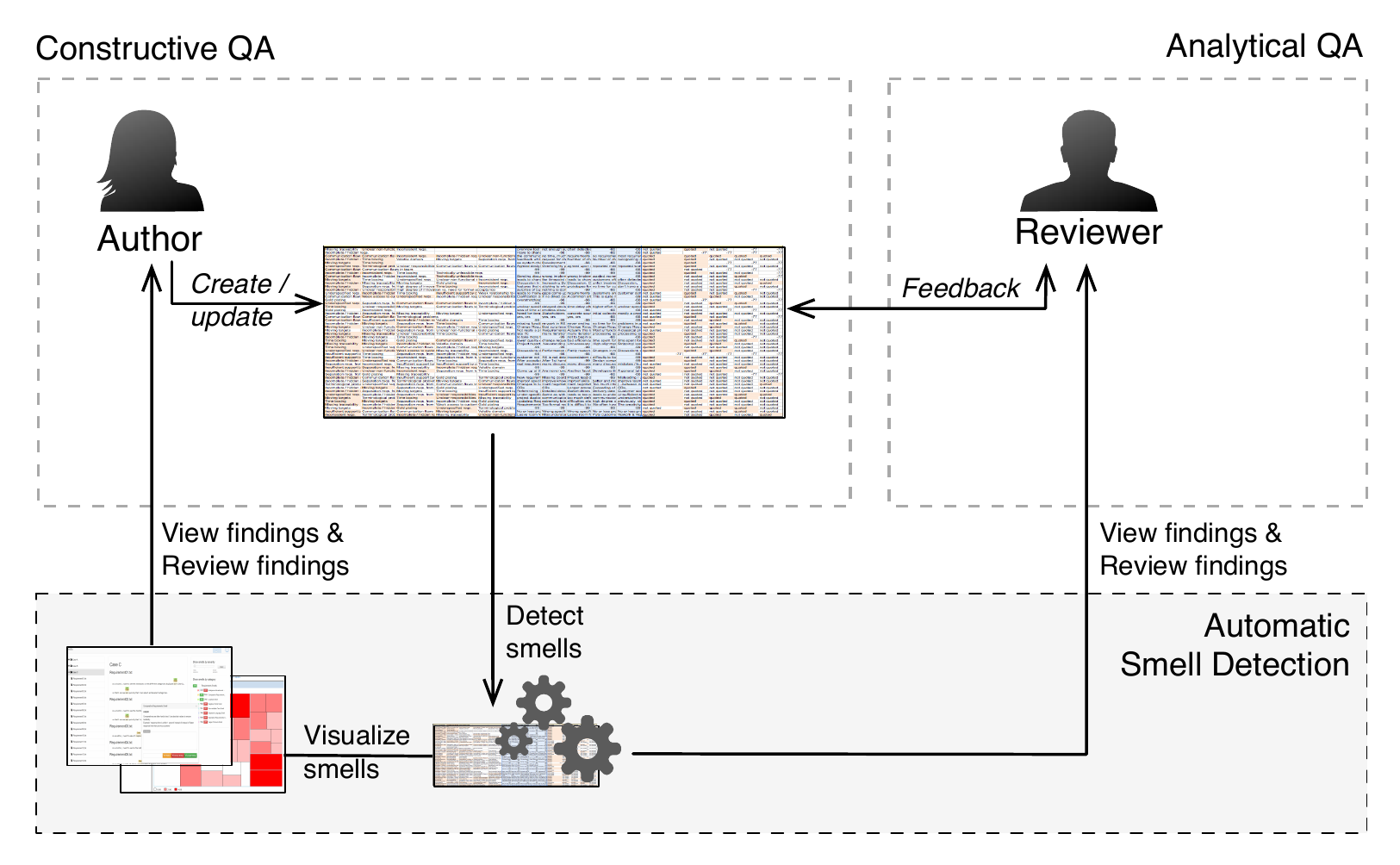}\squeeze
\caption{A suggestion for applying Requirements Smell detection in QA}\squeeze
\label{fig:qa_process}
\end{center}
\end{figure*}

We postulate the applicability of the Requirements Smell detection in the process of both constructive and analytical quality assurance (see Fig.~\ref{fig:qa_process}). 
From the perspective of a constructive quality assurance, authors can use the smell detection to increase their awareness of potential smells in their requirements artifacts and to remove smells before releasing an artifact for, e.g., an inspection. External reviewers in turn, can then use the smell detection to prepare analytical, potentially cost-intensive, quality assurance tasks, such as a Fagan inspection~\cite{fagan2002design}. Such an inspection involves several reviewers and would benefit from making potential smells visible in advance. Iterative inspection approaches are also known as phased inspections, as defined by Knight and Myers~\cite{Knight1993a}.

We furthermore believe that one major advantage is that the scope of our smell detection is not to enforce resolving a potential smell but to increase the awareness of the like and to make transparent later reasoning why certain decisions have been taken. Please note that two different roles (e.g. requirements engineer and QA engineer) can take two different viewpoints on the same smell, respectively its criticality and whether it should be resolved or not. In addition, a finding could be unambiguous to the author, but unclear to the target group of readers (represented by the reviewers). Therefore, one contribution of our tool-supported smell detection is also to actively foster the communication between reviewers and authors and to enable continuous feedback between both roles. For this reason, we enable stakeholders in Smella to comment on detected smells and make explicit whether they need to be resolved or whether and why they have been accepted or rejected.

\section{Evaluation}
\label{sec:Evaluation}

For a better, empirical understanding of smells in requirements artifacts, we conducted an exploratory multi-case study with both industrial and academic cases. We particularly rely on case study research over other techniques, such as controlled experiments, because we want to evaluate our approach in practical settings under realistic conditions. For the design and reporting of the case study, we largely follow the guidelines of Runeson and H\"ost~\cite{Runeson2008}.

\subsection{Case study design}
\label{ssec:cs_design}

Our overall research objective is as follows:\\

\noindent\textbf{Research Objective:} Analyze whether automatic analysis of Requirements Smells helps in requirements artifact quality assurance.\\

To reach this aim, we formulate four research questions (RQ). In the following, we introduce those research questions, the procedures for the case and subjects selection, the data collection and analysis, and the validity procedures.

\subsubsection{Research questions}
%To achieve our overall research objective, we formulate four research questions. 

\noindent\textbf{RQ~1: How many smells are present in requirements artifacts?} %(and ratios)
 To see if the automatic detection of smells in requirements artifacts could help in QA, we first need to verify that Requirements Smells exist in the real world. The answer to this question fosters the understanding how widespread the smells under analysis are in industrial and academic requirements artifacts. 
 %\todo[inline]{@Henning: rephrase: automatically detectable, relevant}

\noindent\textbf{RQ~2: How many of these smells are relevant?} 
Not only the number of detected smells is important. If many of the detected smells are false positives and not relevant for the requirements engineers and developers, it would hinder QA more than it would help. As relevancy is a rather broad concept, we break down RQ~2 into two sub-questions. 

\begingroup
\leftskip2em
\noindent\textbf{RQ~2.1: How accurate is the smell detection?} % (internal review to the researchers)
 The first sub-question looks at the more technical view on relevance. We want to find false positives and false negatives to determine the precision and recall of the analysis in terms of correct detection of the defined smell. 

\noindent\textbf{RQ~2.2: Which of these smells are practically relevant in which context?} %(external review of a preselected subset with practitioners)
 This second sub-question is concerned with practical relevance. We investigate whether practitioners would react and change the requirement when confronted with the findings.

\endgroup

\noindent\textbf{RQ~3: Which requirements quality defects can be detected with smells?} %quality
%(relation to IEEE criteria with Stuttgart)
 After we understood how relevant the analyzed Requirements Smells are, we want to understand their relation to existing quality defects in requirements artifacts. Hence, we need to check whether, and if so, which defects in requirements artifacts correspond to smells, as we understand smell findings as indicators for defects. %\todo{(HF) passt das so? Oder sollen wir noch schreiben, dass wir die "relation to the existing method for QA, reviews" suchen? DM: Denke das passt so, alles andere koennte unnoetige zusaetzliche Erwartungen / Faesser i.S.v. related work aufmachen.} 

\noindent\textbf{RQ~4: How could smells help in the QA process?} %(qualitatively with practitioners)
 Finally, we collect general feedback from practitioners whether (and how) smell detection could be a useful addition to QA for requirements artifacts and whether as well as how they would integrate the smell detection into their QA process.

\subsubsection{Case and subjects selection}

Our case and subject selection is opportunistic but in a way that maximizes variation and, hence, evaluates the smell detection in very different contexts. This is particularly important for investigating requirements artifacts under realistic conditions, also due to the large variation in how these artifacts manifest themselves in practice. A prerequisite for our selection is the access to the necessary data. To get a reasonable quantitative analysis of the number of smells (RQ~1) and qualitative analysis of the relation of smells and defects (RQ~3), we complement our three industrial cases with a case in an academic setting. There, various student teams are asked to provide software with a certain set of (identical) functionality for a customer as part of a practical course. This is also a realistic setting but provides us with a higher number of specifications and reviews than in the industrial cases.
%additionally gives us the possibility to cover a broader spectrum of specifications.%we analyzed specifications and reviews written by students to answer RQ~3 as this gave us the possibility to compare the results to defects found in reviews covering a broader spectrum of specifications.
%\todo{(HF) Why students for RQ~3? DM: Passt das so? (Noch offen im Rebuttal)}

%Finally, for a reasonable quantitative analysis of the relation of smells and defects, we added cases from student projects. In these projects, student teams are asked to provide software with a certain set of (identical) functionality for a customer. 
We will refer to the subjects of the industrial cases as \emph{practitioners} and we will call the latter subjects \emph{students}.

\subsubsection{Data collection procedure}
\label{sec:data_collection}
We used a 6-step procedure to collect the data necessary for answering the research questions. %As we could not collect all data from all cases, we performed not all steps for all cases (see Table~\ref{tbl:tab:studyObjectsVSResearchQuestions).

\begin{enumerate}
\item \emph{Collect requirements artifact(s) for each case.} We retrieved the requirements artifacts to be analyzed in each case. 
	For one case, the requirements were stored in Microsoft Word Documents. For the other cases, this involved extracting the requirements from other systems, either a proprietary requirements management tool (resulting in a list of html files), or the online task management system JIRA, which led to a set of comma-separated values files. For the student projects, the students handed in their final artifacts either as a single PDF or as a PDF with the general artifact and another PDF with the use cases. Where authors explicitly structured requirements in numbered requirements, user stories or use cases, we counted these artifacts.
\item \emph{Run the smell detection via Smella.} We applied our detection tool as introduced in Sect.~\ref{sec:Smella} on the given requirements artifacts, which generated a list of smells per artifact.
\item \emph{Classify false positives.}\label{step:precision} For all cases in which we wanted to present our results to practitioners, we reviewed each detected finding. In pairs of researchers, we classified the findings as either true or false positive. We classified a finding as false positive if the finding was not an instance of the smell, e.g.\ because the results of the linguistic analysis was incorrect.\footnote{For example, if the linguistic analysis incorrectly classified the word \emph{provider} in the sentence \emph{``As a provider, I want [\ldots]''} as a comparative adjective.} For artifacts containing more than 10 findings of a smell, we only inspected a set of 10 random findings (of that smell) per artifact. The same holds for Case D, where we inspected 10 random findings of each category for the whole case.
\item \emph{Inspect documents for false negatives.}\label{step:recall} To calculate the recall of the smell detection, for each case we randomly selected one artifact that a pair of researchers inspected for false negatives. To ease the manual inspection, we grouped the smells \texttt{Subjective Language}, \texttt{Ambiguous Adverbs and Adjectives}, \texttt{Loopholes}, \texttt{Non-verifiable Terms} (as \texttt{Ambiguity-related smells}). We classified whether a finding is a true or false negative based on the same conditions as in the previous step. 
\sloppy%this prevents overfull hlines in texttt names of smells

One common cause for false negatives for dictionary-based smells can be that an ambiguous phrase is not part of the dictionary. Since we developed the dictionaries based on existing dictionaries, such as the standard, these dictionaries are not yet complete and must be further developed. However, since this is an issue that is not a problem of the smell detection approach in general, but rather a configuration task, we did not take these findings into consideration for the recall.

\item \emph{Get rating by practitioners.} \label{step:rating}
%We took the list of findings per study object of the cases with practitioners and, first, removed the false positives found in the previous step. 
	We selected a subset of the true positive findings so that we cover all smells with a minimum of two findings per smell as far as the artifacts allowed. When we found repeating or similar findings, e.g. multiple similar sentences with the same smell, we also included one of these findings into the set. %\todo{(HF): Oder sollen wir diesen Satz weglassen? Hat das einen Einfluss auf die Studie? DM: Wuerde ich der Vollstaendigkeit drin lassen}
	
	We presented this subset to the practitioners and interviewed them, finding by finding, through three closed questions (see  also Table~\ref{tbl:findings_examples}): Q1:~Would you consider this smell as relevant? Q2:~Have you been aware of this finding before? Q3:~Would you resolve the finding? Of these, the former two must be answered with \emph{yes} or \emph{no}. For the last question, we also needed to take the criticality into account. Therefore, in case practitioners answered that they would resolve a finding, we also asked whether they would resolve it immediately, in a short time (i.e.\ within this project iteration) or in a long time (e.g.\ if it happens again). In addition to these three questions, we took notes of qualitative feedback, such as discussions.
\item \emph{Interview practitioners.} \label{step:interview} In addition to the ratings, we performed open interviews with practitioners about their experience with the smell detection and how they might include it in their quality assurance process. We took notes of the answers. 
\item \emph{Get review results from students.} Lastly, the students performed reviews of the artifacts of other student teams. They documented and classified found problems according to a checklist (see Table~\ref{tbl:checklist}) without awareness of the smell findings in their artifacts. We then collected the review reports from the students.
%\todo[inline]{@Stefan: add guidelines/review check list}
%\todo{Koennen wir diesen Punkt mit Punkt 4 verschmelzen, letztendlich haben wir beides in einem Atemzug gemacht?}
\end{enumerate}

\subsubsection{Analysis procedure}

We structure our analysis procedure into seven steps. Each step leads to the results necessary for answering one of our
research questions.

\begin{enumerate}
\item \emph{Calculate ratios of findings per artifact}. To understand whether smells are a common issue in requirements
	artifacts, we compared the quantitative summaries of smells in the various artifacts and domains. To enable a comparison between different types of requirement artifacts, we used the number of words in each artifact as a measure of size. Hence, we finally reported the ratio of findings per 1000 words for each smell and all smells in total. This provided answers for RQ~1.
\item \emph{Calculate ratios of findings for parts of user stories.} In one case, we had a common structure of the requirements, because they were formulated as user stories. To get a deeper insight into the distribution of smells and findings, we calculated
	the ratios of findings per 1000  words for each part. We divided the user stories into the parts \emph{role} (``As a\ldots''), \emph{feature}
	(``I want to\ldots'') and \emph{reason} (``so that\ldots'') using regular expressions. We counted the words and findings in each part. 
%	but 290 of US have no reason part, reason is usually shorter, so we need to normalise by number of words in each part (RQ~1)
	This provided further insights into the answer for RQ~1.
\item \emph{Calculate ratios of false positives.} After a rough overview obtained under the umbrella of RQ~1 describing the number of findings for each smell of the varying artifacts, we wanted to better
	understand the smell's relevance. The first step was to calculate the ratios of false positive as we classified them in Step~\ref{step:precision} of the data
	collection. We reported false positive rates overall and for each smell. This provides the first part of the answer to RQ~2.1.
\item \emph{Calculate ratios of false negatives.} The precision of a smell detection is tightly coupled with the recall. Therefore, we calculated the ratio of detected smell findings to all existing findings, according to our manual inspection, as described in Step~\ref{step:recall} of the data collection procedure. This provides the second part of the answer to RQ~2.1.
\item \emph{Calculate ratio of irrelevant smells.} We were not only interested in errors in the linguistic analysis but also in
	how relevant the correct analyses were for the practitioners. Hence, we calculated and reported the ratios of findings considered irrelevant by the practitioners. This answers RQ~2.2.
\item \emph{Compare defects from reviews with findings.} From the students, we received review reports for each artifact. As the
	effort to check them all would have been overwhelming, we took a random sample of 20\% of the artifacts. For each of
	the defects detected in the review, we checked if there is a corresponding finding from a smell. This answers RQ~3. %Abstract from defect. dectetablness: yes/no. If no, could it be detectable?
	%&\todo{(HF): Why students?}
\item \emph{Interpret interview notes.} To answer finally RQ~4, we analyze the interview transcripts and code the answers given by the interviewees manually. 
\end{enumerate}

\subsubsection{Validity procedure}
\label{sec:ValidityProcedure}
%\todo[inline]{TODO: Stefan}
 
First, we used peer debriefing in the sense that all data collection and analyses were done by at least two researchers. Analysis results were also checked by all researchers. This researcher triangulation especially increases the internal validity. Furthermore, we kept an audit trail in a Subversion system to capture all changes to documents and analyses.
 
Second, we performed all the classifications of findings into true and false positives in pairs. This already helped to avoid misclassifications. To further check our classifications, we afterwards did an independent re-classification of randomly selected 10\% of the findings and calculated the inter-rater agreement. We discussed to clarify which findings we consider false positives and repeated the classifications until we reached an acceptable agreement. 
The same procedure held for the inspection of artifacts to detect false negatives, which we also conducted in pairs. Furthermore, we also independently re-classified one of the artifacts to understand the inter-rater agreement on the false negatives. Overall, our analysis for false positives and relevance of the findings is also a validity procedure in the sense that we check in RQ~2 the results from RQ~1. 

Third, we discussed with the practitioners what relevance of smells means in the context of the study to avoid misinterpretations. Furthermore, we gave the students review guidelines to give them an indication what quality defects in requirements artifacts might be. Both serve in particular as mitigation to threats to the internal and the construct validity.

Fourth, we performed the analysis of the correspondence between smells and defects with a pair of researchers. This pair derived a classification of the found and not found defects. Both other researchers reviewed the classification, and we improved it iteratively until we reached a joint agreement.

Fifth, we performed member checking by showing our transcriptions and interpretations for RQ~4 to the interviewed practitioners and incorporating feedback.%\todo[inline]{HF: not interpretations!}

Finally, to support the external validity of the results of our study, we aimed at selecting cases with maximum variation in their
domains, sizes, and how they document requirements.

\subsection{Results}
\label{ssec:cs_results}
In the following, we report on the results of our case studies. We first describe the cases and subjects under analysis, before we answer the research questions. We end by evaluating the validity of the cases.

\subsubsection{Case and subjects description}
%\todo[inline]{Why is this here and not directly at Case and Subjects Selection? Actually, I would not consider this as result.}
%We intentionally chose cases varying in size and methodology followed. 
The first three cases contain requirements produced in different industrial contexts: embedded systems in the automotive industry, business information systems for the chemical domain and agile development of web-based systems. While the first two represent rather classical approaches to Requirements Engineering, the third case applies the concept of user stories, as it is popular in agile software development. The fourth case is in an academic background and employs both use cases and textual requirements. Regarding subject selection, for each industrial case we selected practitioners involved in the company, domain and specification. We executed the findings rating (Step~\ref{step:rating}) and the interviews regarding the QA process (Step~\ref{step:interview}) with the same experts, so that their answer in Step~\ref{step:interview} is based on their experience with practical, real examples. In the following, we describe the cases, as well as the experts or students for each case.
Table~\ref{tbl:objects} provides a quantitative overview of the cases.

\paragraph{Case A: Daimler AG}
Daimler AG is a multinational automotive corporation headquartered in Stuttgart, Germany. At Daimler, we analyzed six different requirements artifacts (A1--A6) which were written by various authors. The requirements artifacts describe functionality in different domains of engine control as well as driving information. In this case, requirements are written down in the form of sentences, identified by an ID. The authors are domain experts who are coached on writing requirements.

The requirements artifacts A1--A6 consist of 323 requirements in total (see Table~\ref{tbl:objects}). All of the artifacts of Daimler analyzed in our study were created by domain experts in a pilot phase after a change in the requirements engineering process as part of a software process improvement endeavour. For RQ~2.2., we reviewed 22 findings with an external coach who works as a consultant for requirements engineering and has tightly collaborated with the group for many years. 

\paragraph{Case B: Wacker Chemie AG}
In the second case, we analyzed requirements artifacts of business information systems from Wacker Chemie AG. Wacker is a globally active company working in the chemical sector and headquartered in Munich, Germany. The systems that we analyzed fulfil company-internal purposes, such as systems for access to Wacker buildings or support systems for document management.

We analyzed three Wacker requirements artifacts that were written by five different authors. At Wacker, functional requirements are written as use cases (including fields for \emph{Name}, \emph{Description}, \emph{Role} and \emph{Precondition}) whereas non-functional requirements are described in simple sentences. The artifacts consisted of 53 use cases and 13 numbered requirements (see Table~\ref{tbl:objects}). For the reviews of the findings in RQ~2.2, we selected 18 findings and discussed them with the Chief Software Architect, who also has several years of experience in quality assurance.

\paragraph{Case C: TechDivision}
For the third case, we analyzed the requirements of the agile software engineering company TechDivision GmbH. TechDivision has around 70 employees, working in 3 locations in Germany. They focus mainly on web development, i.e.\ creating product portals and e-commerce solutions for a variety of companies, as well as web consulting, especially focusing on search engine optimizations. Many of their products involve customisation of Magento\footnote{\url{http://www.magento.com}} or Typo3\footnote{\url{http://www.typo3.org}} frameworks.

In their projects, TechDivision follows an agile software development process using either Scrum~\cite{schwaber2011scrum} or Kanban~\cite{anderson2010kanban} methodologies. For their requirements, TechDivision applies user stories~\cite{cohn2004user}, which they write and manage in Atlassian JIRA\footnote{\url{https://atlassian.com/software/jira}}. User stories at TechDivison follow the common Connextra format: \emph{As a [Role], I want [Feature], so that [Reason]}. We will also follow this terminology here.

The systems under analysis consist of two online shopping portals, a customer-relationship system and a content-management system, all of which we cannot name for non-disclosure-agreement reasons. In total, we analyzed over 1,000 user stories containing roughly 28,000 words.  For RQ~2.2, we met with an experienced Scrum Master and a long-term developer, who have worked on several projects for TechDivision. 

\paragraph{Case D: University of Stuttgart}
The requirements of Case D were created by 52 groups of three 2nd-year students each during a compulsory practical course in the software engineering programme at the University of Stuttgart. We removed one artifact, because it was incorrectly encoded, thus resulting in 51 requirements artifacts for this analysis.

\begin{figure}[htb]
\begin{center}
\includegraphics[width=.6\columnwidth]{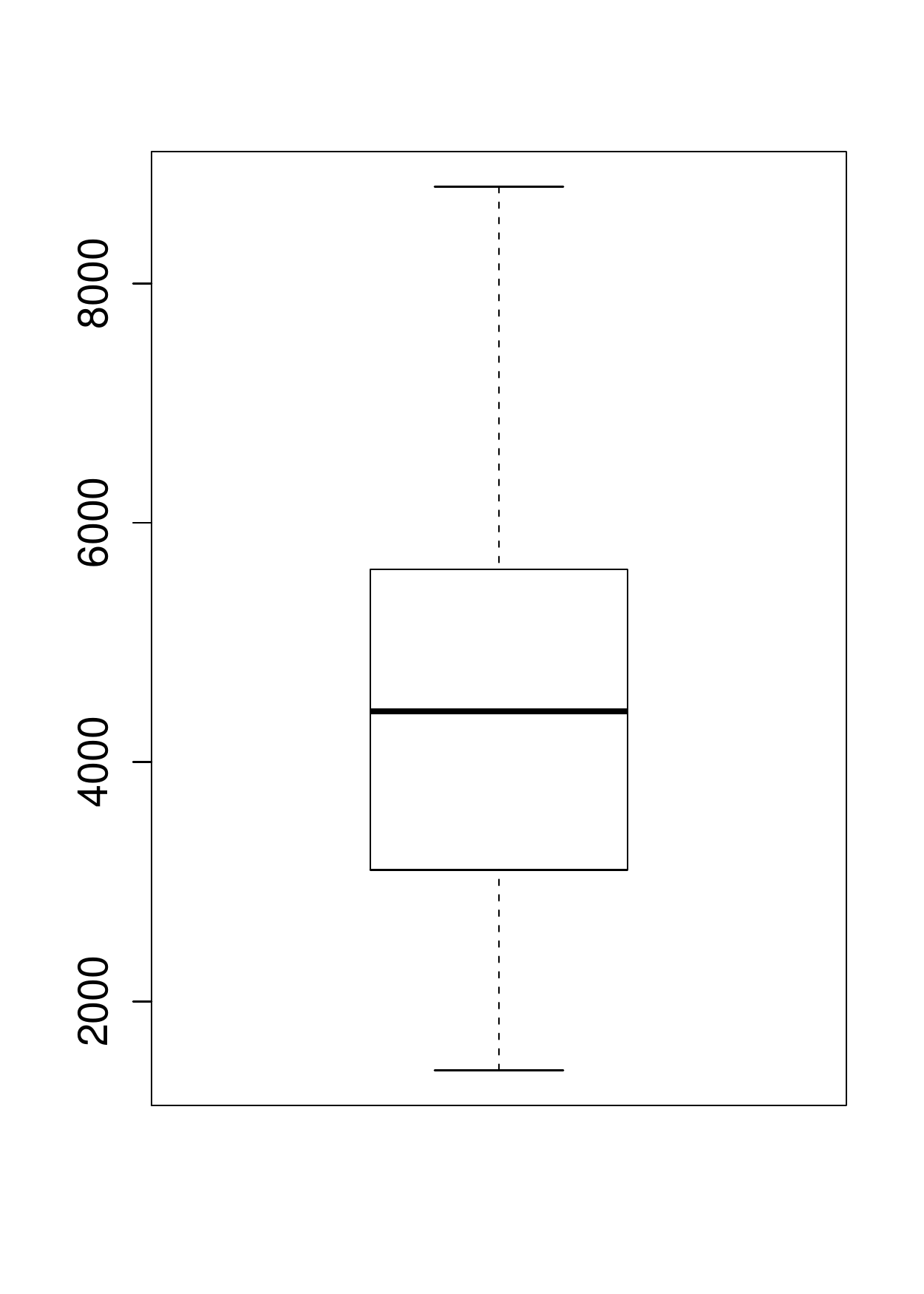}
\caption{Variation of size of requirements artifacts in Case D in words}
\label{fig:stuttgart_size_words}
\end{center}
\end{figure}

The resulting requirements artifacts differ vastly in style; hence, we were unable to count them in terms of requirements, but instead only counted the structured use cases as provided by the authors, and quantified the artifacts by word size. 
The average size of a requirements artifact was 4,471 words (min: 1,425, max: 8,807, see Fig.~\ref{fig:stuttgart_size_words}) and contained 19 use cases (min: 6, max: 39), thus creating a set of artifacts of nearly a quarter of a million words, including more than 950 use cases.

\begin{table*}[hbt]
\caption{Study objects}\label{tbl:objects}%\scriptsize
\centering
\begin{tabular}{l l l l l l l}\toprule
Artifact & Topic & \rot{Size in Words} & \rot{\# Requirements} & \rot{\# Use Cases} & \rot{\# User Stories} \\
\midrule
A1 & Adaptive valve control & 1896 & 91 & \\
A2 & Exhaust control & 2244 & 72 & \\
A3 & Driving information & 199 & 12 & \\
A4 & Engine startup control & 975 & 44 & \\
A5 & Engine control & 524 & 49 & \\
A6 & Powertrain communication & 1100 & 55 &\\\midrule
Sum Daimler &  & 6938 & 323 & \\
\midrule\midrule
B1 & Management of access control & 2093 & 9 & 18 \\
B2 & Event notification & 1015 & 3 & 19\\
B3 & Document management & 458 & 1 & 16 \\\midrule
Sum Wacker & & 3566 & 13 & 53 \\ 
\midrule\midrule
C1 & Webshop for fashion articles & 5226 & & & 168 \\
C2 & CMS in transportation domain & 2742 & & & 123\\
C3 & CRM system & 6863 & & & 230\\
C4 & Webshop for hardware articles & 13124 & & & 561\\
\midrule
Sum TechDivision & & 27955 & & & 1082 \\
\midrule\midrule
Avg Stuttgart &  & 4470 & & 18.9 &\\
\midrule
Sum Stuttgart & & 227973 & & 966 &\\
\midrule\midrule
Sum over all & & 266432 & 336 & 53 & 1082 \\
\bottomrule
\end{tabular}
\end{table*}
%A1 Komponentenfunktionen 
%A2 NDAGR-Lernfunktion
%A3 Reiserechner
%A4 Starteransteuerung 
%A5 Verbrennungsregelung 
%A6 Vernetzung Powertrain
%
%B1 Fremdfirmenanmeldung 
%B2 osmar.net 
%B3 pod
%
% C1 ANITA
% C2 EUROTOURS
% C3 Oelcheck
% C4 Pro Aurum

For practical reasons, we could not evaluate each research question in each case: For example, RQ~3 depends on the existence of reviews with documented results, which is often not existent in practice. Furthermore, depending the answers of RQ~4 on the potentially less experienced students from Case D would introduce a threat to the validity of our evaluation. Table~\ref{tab:studyObjectsVSResearchQuestions} shows the mapping between research questions and study objects.
The interviews for RQ~2.2 and RQ~4 lasted 60 minutes for each Case A and B and 120 minutes for Case C.

\begin{table}[htb]
\begin{center}
\caption{Study objects usage in research questions}
\label{tab:studyObjectsVSResearchQuestions}
\begin{tabular}{l c c c c c c} \toprule
 Case  & \rot{RQ 1: Distribution} & \rot{RQ 2.1: Precision} & \rot{RQ 2.1: Recall} & \rot{RQ 2.2: Relevance} & \rot{RQ 3: Defect Types} & \rot{RQ 4: QA Process} \\ \midrule
 A: Daimler & \checkmark & \checkmark & \checkmark &   &  &  \checkmark \\
 B: Wacker & \checkmark & \checkmark & \checkmark &   &  & \checkmark \\ 
 C: TechDivision & \checkmark & \checkmark & \checkmark & \checkmark &  & \checkmark \\ 
 D: Univ.\ of Stuttgart & \checkmark & \checkmark &  \checkmark  &  & \checkmark &  \\ 
 \bottomrule
\end{tabular}
\end{center}
\end{table}

\subsubsection{RQ 1: How many Requirements Smells are present in the artifacts?}
\label{sec:rq1}

\begin{sidewaystable*}[htbp]
\caption{Quantitative summary of smell findings}\label{tbl:smells_all}\scriptsize
\centering
\begin{tabular}{@{}l p{1cm}*{9}{p{0.6cm}p{0.6cm}}@{}}
\toprule
\multirow{2}{*}{Case} & \multirow{2}{1.2cm}{Num Words} & \multicolumn{2}{p{1.2cm}}{All Smells} & \multicolumn{2}{p{1.2cm}}{Subjective Language Smell} & \multicolumn{2}{p{1.2cm}}{Loophole Smell} & \multicolumn{2}{p{1.2cm}}{Vague Pronouns Smell} & \multicolumn{2}{p{1.2cm}}{Superlatives Smell} & \multicolumn{2}{p{1.2cm}}{Negative Words Smell} & \multicolumn{2}{p{1.2cm}}{Comparatives Smell} & \multicolumn{2}{p{1.2cm}}{Non-verifiables Smell} & \multicolumn{2}{p{1.2cm}}{Ambiguous A \& A Smell} \\
 &  & abs & rel & abs & rel & abs & rel & abs & rel & abs & rel & abs & rel & abs & rel & abs & rel & abs & rel \\ \midrule
A1 & 1896 & 45 & 23.7 & 4 & 2.11 & 2 & 1.05 & 13 & 6.86 & 7 & 3.69 & 11 & 5& 7 & 3.69 & 0 & 0& 1 & 0.53 \\
A2 & 2244 & 52 & 23.2 & 6 & 2.67 & 3 & 1.34 & 20 & 8.91 & 1 & 0.45 & 14 & 6.24 & 5 & 2.23 & 2 & 0.89 & 1 & 0.45 \\
A3 & 199 & 5 & 25.1 & 0 & 0& 0 & 0& 3 & 15.08 & 0 & 0& 2 & 10.05 & 0 & 0& 0 & 0& 0 & 0\\
A4 & 975 & 29 & 29.7 & 3 & 3.08 & 1 & 1.03 & 15 & 15.38 & 0 & 0& 8 & 8.21 & 1 & 1.03 & 1 & 1.03 & 0 & 0\\
A5 & 524 & 20 & 38.2 & 0 & 0& 0 & 0& 14 & 26.72 & 0 & 0& 5 & 9.54 & 0 & 0& 1 & 1.91 & 0 & 0\\
A6 & 1100 & 32 & 29.1 & 0 & 0& 0 & 0& 8 & 7.27 & 0 & 0& 13 & 11.82 & 7 & 6.36 & 4 & 3.64 & 0 & 0\\
\midrule
Sum Daimler & 6938 & 183 & 26.4 & 13 & 1.87 & 6 & 0.86 & 73 & 10.52 & 8 & 1.15 & 53 & 7.64 & 20 & 2.88 & 8 & 1.15 & 2 & 0.29 \\
\midrule
\midrule
B1 & 2093 & 90 & 43& 5 & 2.39 & 11 & 5.26 & 40 & 19.11 & 6 & 2.87 & 20 & 9.56 & 7 & 3.34 & 1 & 0.48 & 0 & 0\\
B2 & 1015 & 28 & 27.6 & 2 & 1.97 & 1 & 0.99 & 13 & 12.81 & 0 & 0& 3 & 2.96 & 9 & 8.87 & 0 & 0& 0 & 0\\
B3 & 458 & 31 & 67.7 & 0 & 0& 19 & 41.48 & 9 & 19.65 & 1 & 2.18 & 0 & 0& 1 & 2.18 & 0 & 0& 1 & 2.18 \\
\midrule
Sum Wacker & 3566 & 149 & 41.8 & 7 & 1.96 & 31 & 8.69 & 62 & 17.39 & 7 & 1.96 & 23 & 6.45 & 17 & 4.77 & 1 & 0.28 & 1 & 0.28 \\
\midrule
\midrule
C1 & 5226 & 229 & 43.8 & 48 & 9.18 & 5 & 0.96 & 104 & 19& 3 & 0.57 & 29 & 5.55 & 36 & 6.89 & 1 & 0.19 & 3 & 0.57 \\
C2 & 2742 & 120 & 43.8 & 11 & 4.01 & 7 & 2.55 & 62 & 22.61 & 3 & 1.09 & 13 & 4.74 & 24 & 8.75 & 0 & 0& 0 & 0\\
C3 & 6863 & 233 & 34& 30 & 4.37 & 14 & 2.04 & 105 & 15& 6 & 0.87 & 31 & 4.52 & 45 & 6.56 & 1 & 0.15 & 1 & 0.15 \\
C4 & 13124 & 572 & 43.6 & 35 & 2.67 & 16 & 1.22 & 339 & 25.83 & 11 & 0.84 & 101 & 7& 49 & 3.73 & 9 & 0.69 & 12 & 0.914 \\
\midrule
Sum TechDivision & 27955 & 1154 & 41.3 & 124 & 4.44 & 42 & 1& 610 & 21.82 & 23 & 0.82 & 174 & 6.22 & 154 & 5.51 & 11 & 0.39 & 16 & 0.57 \\
\midrule
\midrule
Mean Stuttgart & 4470 & 198.5 & 44.4 & 6.45 & 1.44 & 19.65 & 4& 117.37 & 26.26 & 5.12 & 1.14 & 27.59 & 6.17 & 16.63 & 3.72 & 4.71 & 1.05 & 0.96 & 0.21 \\
\midrule
Sum Stuttgart & 227973 & 10122 & 44.4 & 329 & 1.44 & 1002 & 4& 5986 & 26.26 & 261 & 1.14 & 1407 & 6.17 & 848 & 3.72 & 240 & 1.05 & 49 & 0.21 \\
\midrule
\midrule
Over all & 266432 & 11608 & 43.6 & 473 & 1.78 & 1081 & 4.06 & 6731 & 25.26 & 299 & 1.12 & 1657 & 6.22 & 1039 & 3& 260 & 0.98 & 68 & 0.26 \\ \bottomrule
\end{tabular}
\end{sidewaystable*}

Under this research question, we quantify the number of findings that appear in requirements. Table~\ref{tbl:smells_all} shows the number of findings for each case, each requirements artifact and each smell and also puts these numbers in relation to the size of the artifact. We analyzed requirements of the size of more than 250k words, on which the smell detection produced in total more than 11k findings, thus revealing roughly 44 findings per thousand words. 

\begin{sidewaysfigure*}[htbp]
\begin{center}
\includegraphics[width=1\columnwidth]{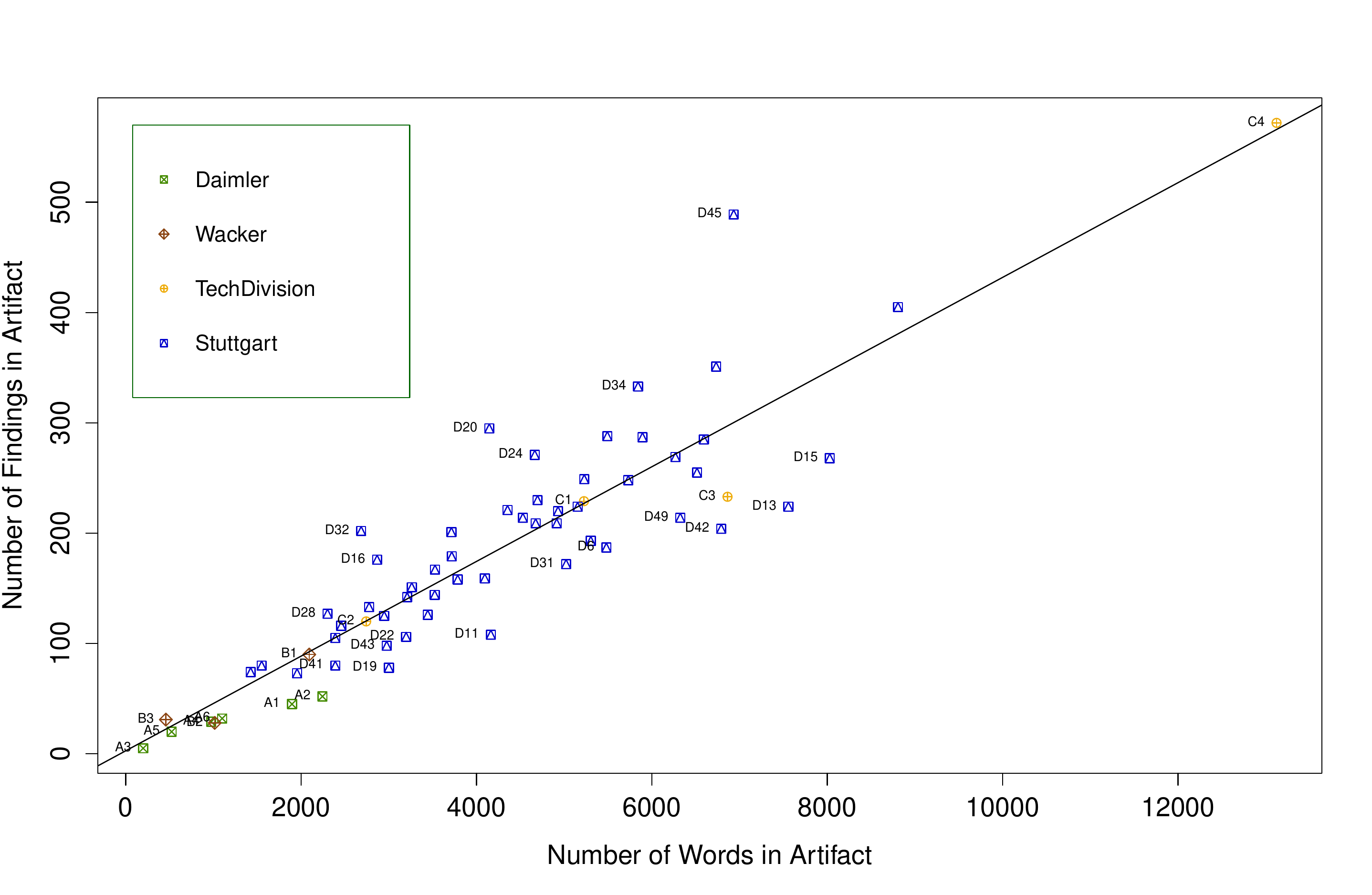}
\caption{Number of findings strongly correlates with size of artifact (for readability reasons, for the Stuttgart cases (blue) only IDs of less correlating artifacts are displayed).}
\label{fig:sizeVsSmells}
\end{center}
\end{sidewaysfigure*}

Table~\ref{tbl:smells_all} shows that all requirements artifacts contain findings of Requirements Smells. They vary from 5 findings for the smallest\footnote{in terms of total number of words} case (A3) up to 572 for the largest case (C4). The number of findings strongly correlates with the size of the artifact (see Fig.~\ref{fig:sizeVsSmells}, Spearman correlation of 0.9). Hence, in the remainder, we normalize the number of findings by the size of the artifact. 

\begin{figure}[htb]
\begin{center}
\includegraphics[width=.6\columnwidth]{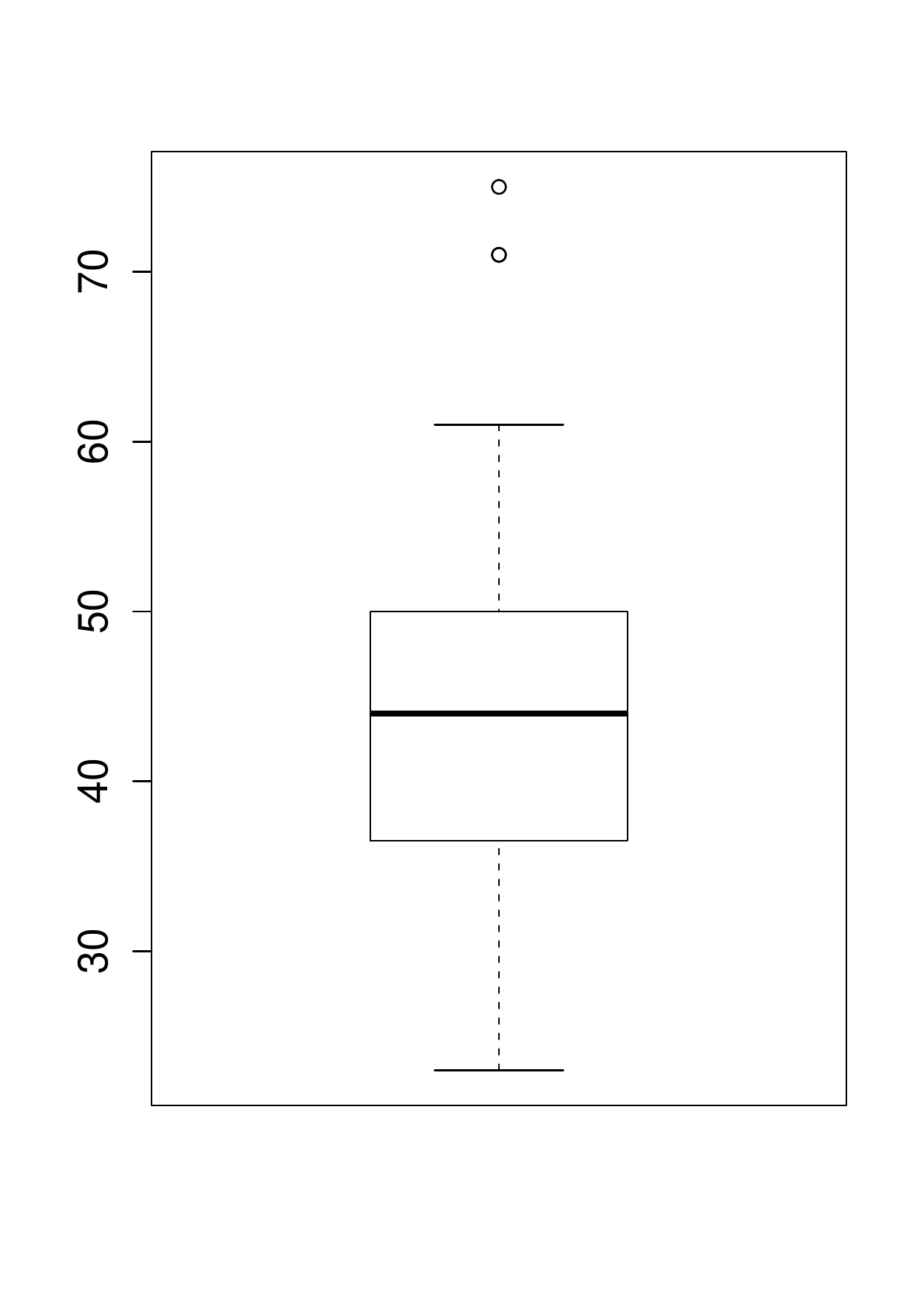}
\caption{Number of findings per 1,000 words in Case D}
\label{fig:stuttgart_all_smells_1000}
\end{center}
\end{figure}

The artifacts of Daimler have an average of 26 findings per thousand words, in contrast to 41 for both Wacker and TechDivision and 43 for the artifacts produced by the students. 
Best to analyze the variance within a requirements artifact seems Case D, in which multiple teams had a similar background and project size. Fig.~\ref{fig:stuttgart_all_smells_1000} shows the variance between the artifacts of Case D with an average of 44 findings, a minimum of 26 findings (D11) and a maximum of 75 findings (D32) per 1,000 words. 

\begin{sidewaysfigure*}[htbp]
\begin{center}
\includegraphics[width=1\columnwidth]{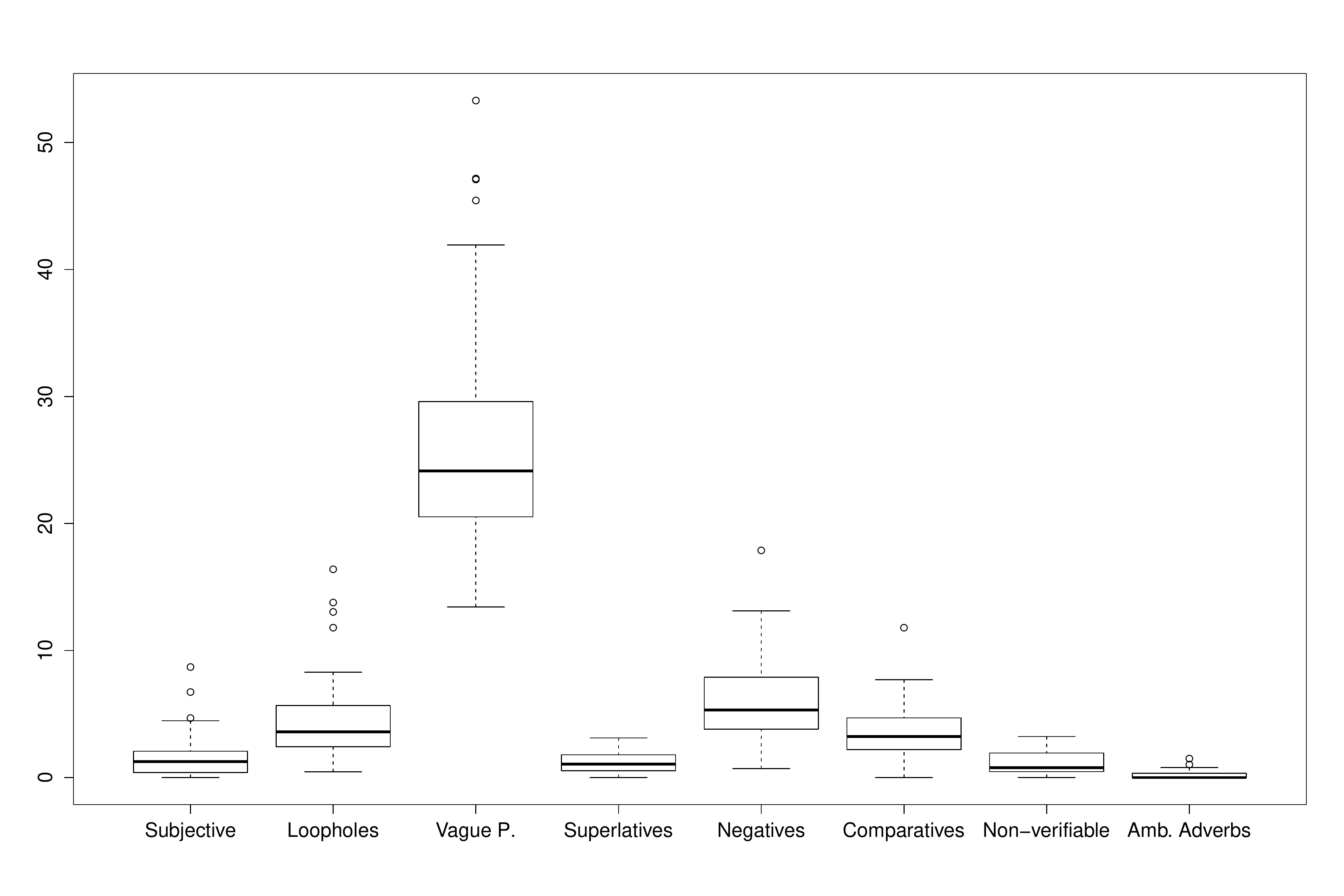}
\caption{Variation of smells per 1,000 words in Case D}
\label{fig:stuttgart_size}
\end{center}
\end{sidewaysfigure*}

When inspecting the different Requirements Smells, we can see that the most common smells are \texttt{vague pronouns} with 25 findings per 1,000 words, followed by the \texttt{negative words} smell with 6 findings and the \texttt{loophole} smell with 4 findings. The least often smells are \texttt{non-verifiable terms} with 1 finding per 1,000 words, and \texttt{ambiguous adverbs and adjectives} with 0.25 findings per 1,000 words. In fact, the most common smell, \texttt{vague pronouns}, appears 100 times more often than the \texttt{ambiguous adverbs and adjectives}. To analyze the variance in depth, we again take the students' artifacts for reference. Fig.~\ref{fig:stuttgart_size} shows the relative number of findings across the projects.

\paragraph{Interpretation}
We interpret the quantitative overview along three variables: projects, contexts and the different Requirements Smells.

\begin{description}
\item[Projects]
When comparing at project level, we see that Cases A1--A6 (with outlier A5) and C1--C4 (with outlier C3) show quite similar numbers. In contrast B1 to B3 vary between 28 and 68 findings per 1,000 words.  When looking into the most extreme outliers B3 and D32, we see a systematic error that creates a large number of findings: Both projects repeatedly explain what the system \emph{should}\footnote{\emph{Soll} is a German modal verb that is less strict than an English \emph{must}.} do instead of what it \emph{must} do. 16 of 19 \texttt{loopholes} findings in B3 and 29 of 37 \texttt{loophole} findings in D32 root from this problem. This can lead to difficult issues in contracting as requirements that are phrased with a \emph{should} are commonly understood as optional (see e.g.~RFC2119~\cite{Bradner1997} for a detailed explanation).

Hence, we could see a surprising consistency in two of three industrial case studies. The Wacker data varies, so does the students case. In both cases, the negative extremes point at issues that potentially have expensive consequences. 

\item[Context]
The four cases differ strongly in their context: They write down requirements in different forms, vary in their software development methodology and  also produce software for different domains. When comparing the findings at the domain level, we see that Daimler artifacts with an average of 26 findings per thousand words contain less findings than both Wacker and TechDivision with 41 findings and the artifacts produced by the students with 43 findings.

Our partners reported that there have been trainings for the authors of the cases A1--A6 recently, which could explain the difference. Another reason could be the strong focus that the automotive domain puts on requirements and requirements quality in contrast to the other domains. Lastly, also the strict process in this domain could be a reason for this striking difference of the Daimler requirements. Unsurprisingly, the students' requirements form the lower end of the scale, yet not by much.

\item[Requirements Smells]
When comparing the eight smells, we see a strong variance between the number of findings, both in absolute as well as relative values. A qualitative inspection indicates reasons for the most occurring smells. First, the smell detection for \texttt{vague pronouns} finds all substituting pronouns in the requirements. Especially in German, in many sentences the reference of the pronoun can sometimes be derived from gender and grammatical case of the word, thus correctly detecting pronouns, but not \emph{vague} pronouns. RQ~2.1 quantifies this issue. Second, the most common indication for \texttt{loophole} findings is the aforementioned use of the word \emph{should}. We discuss this case in-depth with practitioners in RQ~2.2. Third, we will also inspect reasons for the high number of negative words findings in RQ~2.1 and RQ~2.2.

%outlier bei vague: viele personalpronomen: er (der benutzer)/sich (e.g. oeffnet sich ein fenster) /wir/...
\end{description}

\paragraph{Answer to RQ 1}
The number of findings in requirements artifacts strongly correlates with the size of the artifact. There are roughly 44 findings per 1,000 words and some contexts show a striking similarity in the number of findings for their artifacts. In our cases, the automotive requirements had a lower number of findings whereas student artifacts contained a higher number of findings relative to the size of the artifacts. The most common findings are for the smells \texttt{loopholes} and \texttt{vague pronouns}.

\subsubsection{RQ 2.1: How accurate is the smell detection?}
To understand the capabilities of the smell detection, we need to understand precision as metric indicating how many of the detected findings are correct, as well as recall as a metric indicating how many of the correct findings are detected. 

\paragraph{Precision}
To understand to which extent the numbers of findings for certain smells in RQ~1 are caused by the detection mechanism, we inspected a random sample of 616 findings by taking equivalent sets of findings from each project and manually classifying whether the finding fulfills the smell definition. We could not inspect the same number of findings of each smell for each project, because some projects only had few or even no findings of a certain smell (see number of findings per project in Table~\ref{tbl:smells_all}).

\begin{figure*}[htbp]
\begin{center}
\includegraphics[width=.7\textwidth]{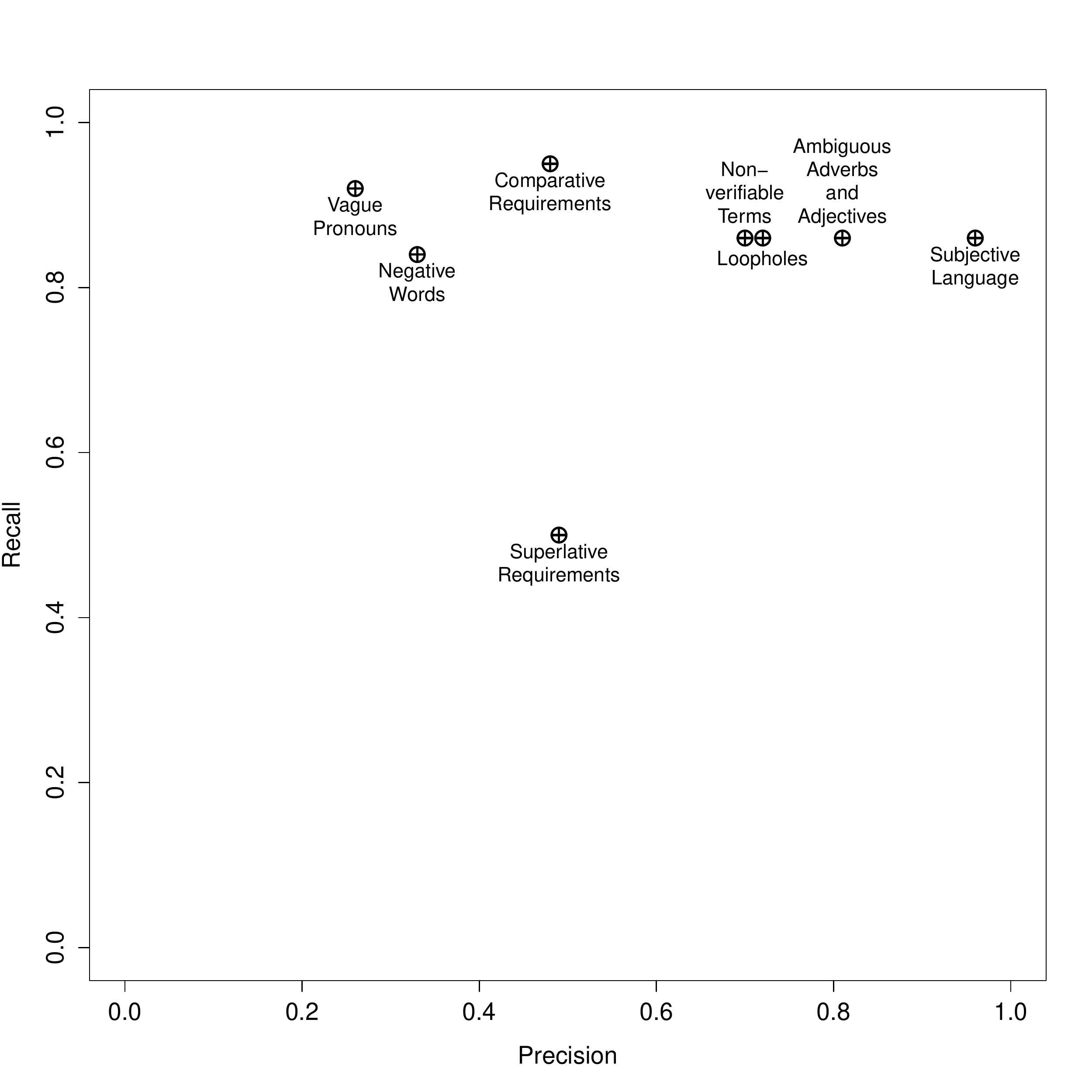}
\caption{Precision and recall of the discussed smell detection approaches.}
\label{fig:precVsRecall}
\end{center}
\end{figure*}

Table~\ref{tbl:precision} and Fig.~\ref{fig:precVsRecall} show the summary of this analysis: The precision of the detection of the \texttt{subjective language} smell revealed only three false positives in total, thus leading to a precision of 0.96. %This means that virtually every finding of this smell produced by the tool is correct. 
\texttt{Non-verifiable words}, \texttt{loophole}, and \texttt{ambiguous adverbs and adjectives} smells range between 0.70 and 0.81, hence leading to roughly one mistake in four suggestions. \texttt{Compara\-tive} and \texttt{superlative} smells range around 0.5 which would mean that every second finding is correct. At the rear end of the list are the \texttt{negative words} and \texttt{vague pronouns} smells with one correct finding in three to four suggestions. Across all smells, the precision is between 0.48 (over all inspections) and 0.59, if we take the varying number of inspected findings between the smells into account. To understand these numbers, we qualitatively inspected the false positive classifications, revealing the following main reasons for false positives:

\begin{description}
\item[Grammatical errors in real world language.] The first issue that creates false positives is the fact that our study analyzes real world language. Some of the requirements, especially in Case C, contained a number of grammatical flaws as well as dialectal phrases, which lead to wrong results in the automatic morphologic analysis and automatic POS tagging and consequently also to false positives during smell detection. 

\item[Vague pronouns.] The smell detection for \texttt{vague pronouns} showed the lowest precision. In the detection of this smell, we look for substituting pronouns, which are pronouns where the noun is not repeated after the pronoun\footnote{E.g. \emph{The father of these}. vs \emph{The father of these kids.}}, of which we characterize only every fourth finding as a defect. The reason behind this poor performance, besides a number of false positives due to the poor grammar mentioned before, is the comparably large number of grammatical exponents of the German language. In addition to number and three grammatical genders, the German language also has four grammatical cases. Therefore, in various instances of substituting pronouns, there is only one grammatical possibility of what the pronoun could refer to.

\item[Findings in conditions.] A third reason for false positives is that the smell detection, so far, takes very little context into account. For example, the \texttt{comparatives} smell aims at detecting requirements that define properties of the system relative to other systems or circumstances\footnote{As discussed in Sect.~\ref{sec:smells_29148}, the problem of comparatives in requirements is validation: How can we understand whether a system fulfills a requirements if that requirement is stated in a relative instead of an absolute way? What if the system in comparison changes its properties, would this render the requirement suddenly unfulfilled?}. When searching for grammatical comparatives in requirements, roughly 48\% of the cases are of the aforementioned kind. In roughly the same number of cases, however, the comparative describes a condition. For example, if the requirement states that \emph{if the system takes more than 1 second to respond [\ldots]}, the comparison is not against another system or circumstance but against absolute numbers. Therefore, in this case, the comparative does not indicate a problem (one could even argue that this is an indicator for \emph{good} quality).

A similar problem holds for the \texttt{negative phrases} smell: The smell detection aims at revealing statements of what the system should not do. Often, however, the negative is mentioned in conditions. For example, if the requirements express what to do \emph{if the user input is not zero [\ldots]}, the negation relates to a condition and not to a property of the system.

%For smells that are based on the meaning of words, a common case for false positives is that the word's level of ambiguity differs from context to context. For example, 
\end{description}

\paragraph{Recall} When analyzing the accuracy of an automatic detection, we must look not only at precision, but also at recall, i.e.\ the ratio of all detected findings to all defects of a certain type in an artifact. To this end, we  inspected one artifact of each case, in total a set of roughly 16,200 words, and manually identified the findings in each artifact. 
Due to the problems of distinguishing the various ambiguity-related smells, we analyzed the recall of these four smells as if it was one smell, without further differentiation (see Section~\ref{sec:data_collection}).

The manual inspection revealed 200 findings in this artifact sample and an average recall of 0.82. Table~\ref{tbl:recall} and Fig.~\ref{fig:precVsRecall} show the summary of the results: The comparison shows a recall between 0.84 and 0.95 for four of the five investigated smells. The highest recall was achieved by the \texttt{Comparative Requirements Smell}, with 0.95, which means that the smell detection missed one in 20 findings.
The fifth smell, with the lowest recall, is \texttt{Superlative Requirements Smell} with a recall of 0.5. However, this smell is one of the rarest of the smells, as one can also see in the results to RQ 1. Therefore our analysis of the recall of this smell is based on few data points. Hence, we suggest to take the recall of this smell with care, and suggest that future studies should investigate this issue in more depth. 

A further analysis of the false negatives shows that the smell detection missed findings because of imprecisions in the NLP libraries (i.e.\ Stanford NLP~\cite{Toutanova2003} for Lemmatization and POS Tagging and RFTagger~\cite{Schmid2008} for morphologic analysis). For the dictionary-based smells, the lemmatization did not correctly deduce the correct lemma, e.g. it did not understand that a certain word was a plural of a lemma. If only the lemmatized version of the word, i.e.\ the singular form, is in the dictionary, then the smell detector does not correctly identify the smell. In the false negative cases for the \texttt{Comparative} and \texttt{Superlative Requirements Smell}, RFTagger did not correctly classify the inflection.

%am meisten, bestmögliche

\paragraph{Interpretation}
The study revealed that the precision strongly varies between the different smells. Qualitative analysis provided further insights described next. 

We can now explain the high number of findings for vague pronouns in RQ~1. If we assume that a quarter of the findings are correct, the number of findings in this category is closer to the remaining smells. Also, we could see that while there are certain reasons of impreciseness that root from the study objects themselves and are, thus, unavoidable, there is plenty of space for optimization. First, existing techniques from NLP could be applied to improve certain smells, such as the \texttt{vague pronouns}. Second, from the examples we have seen, we would argue that the application of heuristics could heavily improve the precision of existing smell detection techniques. For example, if we exploit the information available from POS tagging, we can find out whether a comparison refers to a number or numerical expression. 

Regarding recall, our analysis shows only a slight variance between the smells, with the only outlier being the \texttt{Superlative Requirements Smell}; however, since this is a very rare smell, this recall is based on only few data points, therefore, we must consider this result with care. 
When inspecting the reasons for false negatives, we found that optimizations could be made through the lemmatizer. Future research in this direction should compare whether the accuracy of lemmatizers as reported in the field of computational linguistics also holds for requirements engineering artifacts. Furthermore, we analyzed requirements in German language where lemmatization is a more difficult problem than in English, since the language makes stronger use of inflections (e.g.\ with cases or gender). Hence, smell detectors based on lemmatization for the English language might work better than the results indicate in our analysis.

In general, the precision and recall are therefore comparable to other approaches with related purposes (see Sect.~\ref{sec:rw}). However, is it sufficient for an application of Requirements Smells in practice? 

First, when looking at precision, we must take into account that the current state of practice consists still of manual work and that the cost for running an automatic analysis is virtually zero. Nevertheless, checking a false positive finding takes effort which an inspector could rather spend in reading the document in more detail.  
However, as we see a high variation in the precision over different smells, we need to discuss these separately. Several of the smells have a precision of 0.7 and higher which is considered acceptable in static code analysis~\cite{bessey2010few}. For other Requirements Smells, the precision is below 0.5. This means that every other finding will be a false positive. This can be critical in the effort spent in vain and annoy a user of the smell detection. Yet, we follow Menzies et al.~\cite{Menzies:2007:PPR:1314037.1314090} that a low precision can be still useful ``When there is little or no cost in checking false alarms.'' In our experience, the cost of checking a finding is often just a few seconds. 

Second, when looking at recall, most of the smell detections reach a recall of more than 80\%. Various publications, most prominently Kiyavitskaya~\cite{Kiyavitskaya2008} and Berry et al.~\cite{Berry2012}, argue that a recall close to 100\% is a basic requirement for any tool for automatic QA in RE. The core argument is that with a lower recall, reviewers stop checking these aspects and consequently miss defects, and that reviewers need to check the complete artifact anyway. However, if taking the example of spell checkers and grammar checks, these are still used on a daily basis, although they are far away from 100\% recall. Therefore, one could consequently also argue that the precision is more important than the recall. 

In any case, whether the reported precision and recall are sufficient in industry needs further research in the future. As mentioned above, it mainly depends on two factors: the required investment versus the gained benefit (similar to the concept of technical debt). For the required investment, we argue that, based on our experience of analyzing the various cases presented here, one can quickly iterate through the detected findings with low investment.
To further support this discussion, the following research question analyzes the aspect of the benefits to practitioners in more detail.
%\todo[inline]{add comment about related work}

%\todo[inline]{Do we need to discuss the recall further, e.g. the 100\%-recall argument?}
%\todo[inline]{I think so but only briefly. (Stefan)}

\paragraph{Answer to RQ 2.1}
As shown in Tables~\ref{tbl:precision} and~\ref{tbl:recall}, and as shown in Fig.~\ref{fig:precVsRecall}, the precision is on average around 59\%, with an average recall of 82\%, but both vary between smells. We consider this reasonable for a task that is usually performed manually. However, this also depends on the relevance of findings to practitioners, which we analyze in RQ~2.2. The study also reveals improvements for future work through the application of deeper NLP.

\begin{table*}[htbp]
\centering
\caption{Precision of smell detection}
\label{tbl:precision}
\begin{tabular}{@{}lllll@{}}
\toprule
Smell & \rot{Findings Inspected} & \rot{Findings Accepted} & \rot{Findings Rejected} & \rot{Precision} \\ \midrule
Subjective Language Smell & 69 & 66 & 3 & 0.96 \\
Ambiguous Adverbs and Adjectives Smell & 21 & 17 & 4 & 0.81  \\
Loophole Smell & 60 & 43 & 17 & 0.72 \\
Non-verifiable Term Smell & 23 & 16 & 7 & 0.70 \\
Superlative Requirements Smell & 39 & 19 & 20 & 0.49 \\
Comparative Requirements Smell & 88 & 42 & 46 & 0.48  \\
Negative Words Smell & 129 & 42 & 87 & 0.33  \\
Vague Pronouns Smell & 187 & 48 & 139 & 0.26  \\ \midrule
Average & 77 & 36.6 & 40.4 & 0.59 \\
Overall & 616 & 293 & 323 & 0.48 \\ \bottomrule
\end{tabular}
\end{table*}

\begin{table}[htbp]
\centering
\caption{Recall of smell detection within sample of 4 artifacts (16,271 words)}
\label{tbl:recall}
\begin{tabular}{@{}llllll@{}}
Smell & \rot{Findings in artifacts} & \rot{Findings identified correctly} & \rot{Recall} \\ \midrule
Ambiguity-related S. & 74 & 64 & 0.86\\
Superlative Requirements S. & 4 & 2 & 0.50 \\
Comparative Requirements S. & 21 & 20 & 0.95 \\
Negative Words S. & 64 & 54 & 0.84 \\
Vague Pronouns S. & 37 & 34 & 0.92 \\ \midrule
Average & 40 & 34.8  & 0.82 \\
Overall & 200 & 174 & 0.87\\ \bottomrule
\end{tabular}
\end{table}

%\begin{figure}
%\begin{tikzpicture}
%\begin{axis}[
%	width=\textwidth,
%	height=5cm,
%    ybar,
%    ymin=0, 
%    ylabel=Precision,
%    ymax=1,
%    xtick=data,
%    x tick label style={font=\footnotesize,align=right,rotate=90},
%    every tick/.style={color=black, line width=0.35pt},
%    xticklabels from table={\precisiondatatable}{smell},
%    clip=true,
%	clip mode=individual
%]
%    \addplot table [x expr=\coordindex, y index=1] {\precisiondatatable};
%    \draw[very thin, draw=black, dashed] ({rel axis cs: 0, 0.59}) -- (current axis.right of origin |- {{rel axis cs: 0, 0.59}});
%	\node[right] at (current axis.right of origin |- {{rel axis cs: 0, 0.59}}) {Average};
%\end{axis}
%\end{tikzpicture}
%\caption{Precision of smell detection}\label{fig:precision}
%\end{figure}

\subsubsection{RQ~2.2: Which of these smells are practically relevant in which context?}
To understand whether the Requirements Smells help detecting relevant problems, we first performed a pre-study, in which we confronted practitioners of Daimler and Wacker with findings. The pre-study, which we reported in Femmer et al.~\cite{femmer2014rapid}, aimed at receiving qualitative and tacit feedback. It showed that Requirements Smells can in fact indicate relevant defects. 

In contrast, in this study we analyze relevance in specific categories by interviewing practitioners at TechDivision on their opinion on the findings in terms of relevance, awareness, and whether these practitioners would resolve the suggested finding. 

\begin{sidewaystable*}[hbtp]
\caption{Exemplary findings; shortened and translated by the authors, findings in bold}\label{tbl:findings_examples}
\centering\scriptsize
\begin{tabular}{l p{14cm} l l p{2.2cm}}
\toprule
         ID & Finding & Relevant? & Aware? & Resolve?  \\
   \midrule
   1 & As a visitor, I want to see the checkboxes in the different categories displayed \textbf{more clearly}, so that I can see more quickly that I can select and deselect categories. & Yes & Yes & Yes in short term\\
   2 & As a visitor, I want to see the checkboxes in the different categories displayed more clearly, so that I can see \textbf{more quickly} that I can select and deselect categories. & No & No & No\\
   3 & As an editor, I want to make it \textbf{simpler} to differentiate between … & No & No & No\\
   4 & As a visitor, I want to see \textbf{further} details, \textbf{e.g.} (...) , so that… & Yes & Yes & Yes immediately\\
   5 & As a customer, I want, if I have a \textbf{larger} number of E-Mails in my mailbox, … & Yes & Yes & Yes immediately\\
   6 & As an editor, I want to make it simpler to differentiate between A and B, therefore, A \textbf{should} be labeled as … & Yes & No & Yes in long term\\
   7 & As a provider, I want that, \textbf{as far as possible}, all fields, are mapped between System A and System B. & Yes & Yes & Yes immediately\\
   8 & As a provider I want the news section to be implemented with an effort \textbf{as low as possible}. & Yes & Yes & No\\
   9 & As a visitor, I do \textbf{not} want to see category X, so that I am \textbf{not} confronted with the issue. & No & No & No\\
   10 & As a visitor of the webpage, I want for \textbf{not} selected categories, the displayed hearts of the score (search results list) to be displayed in such a color, that the score display is \textbf{not} changed and always only hearts of relevant categories are displayed in color. & Yes & Yes & Yes immediately\\
   11 & As an employee, I want that an article, if \textbf{no} price is imported, despite the label 'available' to be \textbf{not} displayed in SYSTEM X, so that the article automatically resumes when a price is imported. & Yes & Yes & Yes immediately\\
   12 & As a user, I want to have the possibility to use custom values for \textbf{minimum} and \textbf{maximum}, so that (…). & No & No & No\\
  13 & As a visitor, I want to have a possibility to browse through previous and next products, so that I can \textbf{quickly} and \textbf{easily} look at multiple product without having to go back to the overview page. & No & No & No\\
  14 & As a visitor, I want to navigate to \textbf{meaningfully} structured categories via the menus. & Yes & Yes & Yes immediately\\
  15 & As a visitor, I want to \textbf{quickly} open the pictures of the website, so that unnecessary waiting is avoided. & Yes & Yes & Yes immediately\\
  16 & As a visitor of the website, I want a \textbf{nicely} designed search-suggest-box when I enter a text and wait. & Yes & Yes & Yes immediately\\
  17 & As a buyer, I want to select from a set of shopping providers (…), so that I can select the \textbf{best} suited shopping provider. & No & No & No\\
  18 & As an editor, I want to have multiple entry points for linking categories, so that the visitor can (…) get an overview of selected brands and categories and \textbf{their} filters. & Yes & No & No\\
  19 & As [OTHER SYSTEM], I want that an order of the status 'Order income' transitions into status 'wait for transmission into [SYSTEM]', so that I do not see the order when indexing open orders and so I do not process the order multiple times (and that \textbf{one} can see the status of the order in the backend properly). & No & No & No\\
  20 & As an editor, I want to know a good way how to transfer news content from [SYSTEM] to [SYSTEM] to be able to efficiently migrate \textbf{everything} at once. & Yes & Yes & No\\
  \bottomrule
\end{tabular}
\end{sidewaystable*}

\paragraph{Quantitative observations}
Table~\ref{tbl:findings_examples} reports the 20 findings that we discussed with TechDivision. In summary, we can see that they considered 65\% of the findings as relevant for their context. Furthermore, they have not been aware of 45\% of the findings. Lastly, they would act on 50\% of the presented findings and on 40\% even immediately.

\paragraph{Qualitative observations (true positives)}
The findings that the tool produces mostly constituted forms of underspecification. For example, in Finding~\#1 (see Table~\ref{tbl:findings_examples}): \emph{"As a searcher, I want to see the checkboxes in the different categories displayed \textbf{more clearly}, so that\ldots"} (for similar examples, see Findings 3, 4, 14, 16, and 20). In this case, as in many of the other examples, the practitioners stated that no developer could implement this story properly. They also recalled various discussions in estimation meetings on what was to be done to complete these types of stories\footnote{Note that discussions can have different objectives, i.e.\ \emph{what} is to be implemented and \emph{how}. For these, \emph{how} to implement a story is the team's task and thus discussions can help finding the best way. In contrast, \emph{what} the product owner wants is outside of the team's scope and therefore should not be a matter of discussion.}. 

In the previous research questions, we have seen that Requirements Smells are able to detect \texttt{loopholes} in requirements, such as the usage of the word \emph{should}. To understand the relevance of this finding in the context of an agile company, we also discussed the \texttt{loophole} in Finding~\#6. When we pointed out the finding, they responded that they considered expressing what the system \emph{should} do in user stories problematic. They considered this defect a low risk, as the developers understood (\emph{"If you are told that you should take out the trash, you understand that it is an imperative."}) and their user stories did never turn out to be of legal relevance. They concluded that they want to avoid this, but it has no immediate urgency in a project situation.

ISO~29148 discusses the use of \texttt{negative statements} (\emph{"capabilities not to be provided"}). In a previous study~\cite{femmer2014rapid} practitioners expressed their reluctance of this criterion. In contrast, in this study, practitioners said they would act upon 2 out of 3 of the negative statements (Findings \#9--11) that we presented to them as they revealed unclear requirements. In one case they even remembered that this led to discussions about the implementation during the sprint.
Table~\ref{tbl:findings_examples} shows many more, similar examples.

%    Wann wurde abgelehnt:
%    - wenn die leute es wissen
%    - falsche hoeflichkeit %"Könntest Du bitte den Müll rausbringen?" ist auch eigentlich im imperativ.
%    - process requirements % mit moeglichst wenig aufwand, appell den aufwand in der umsetzung gering zu halten!
%    - bestimmte teile brauchen weniger qualität: rationale

\paragraph{Qualitative observations (false positives)}
Also interesting are those cases that practitioners considered not relevant in their context or where practitioners said they would not act upon. Summarized, the reasons were the following: 
\begin{description}
\item[Domain and context knowledge:] Some stories that were unclear to outsiders were understandable for someone knowing the system under consideration. For example, in user story \#18 it was unclear to the first and second author what \emph{their} refers to. It was clear, however, to both practitioners with knowledge about the system.
\item[Process requirement:] In Finding \#8, the smell reveals another conspicuous finding: The developer should put as \emph{low effort as possible} into the implementation of this story. In the discussion, the reason for this was that the customer did not want to pay much for this implementation. Thus the story should only be fulfilled if it was possible to be fulfilled cheaply. While the practitioners told us they would not change anything about this story, they agreed that the smell pointed out something that violates common user story practice.
\item[Finding in reason part:] In four cases, the practitioners agreed to the finding but considered it irrelevant as the finding was inside the \emph{reason} part of the user story. This is due to this part of the user story only serving as additional information. This reason part is not used in testing nor is the information directly relevant for implementation. The main purpose is to understand the business value and to indicate the major goal to the team, similar to goals and goal modeling in traditional requirements engineering~\cite{AxelvanLambsweerde}.
\end{description}

\paragraph{Answer to RQ~2.2}
In summary, the practitioners expressed that 65\% of the discussed findings were relevant, as they lead to lengthy discussions and unnecessary iterations in estimation. They also saw the problem of legal binding, but in contrast to the practitioners of Case A and B, they considered these findings less relevant. Due to these results, they expressed their strong interest in exploring smell detection for projects; we will explain the results of this discussion in RQ~4. 

\subsubsection*{Further observations of quality defects in different parts of a user story}

%\todo[inline]{RQ~2.2.1 was not explained.}
%\todo[inline]{Reading it now, I don't think it is a good idea to present it as a new RQ here. Maybe it would be okay to just say we noticed that during the analysis and explored a bit further. (Stefan)}

%\paragraph{RQ~2.2.1: How present are findings in the different parts of a user story?}
We considered especially the last explanation for rejecting findings (finding in reason part of a user story) particularly interesting. We had noticed that the reason part was often written in a rather imprecise way. To be able to quantify this aspect, we automatically split user stories according to the language patterns and quantified the distribution of words as well as findings over the different parts of user stories. 

\begin{sidewaystable*}[htbp]
\caption{Findings in different parts of user stories (T=total, Ro=role, F=feature, Re=reason)}
\label{tbl:findings_in_parts}
\begin{tabular}{@{}lllllllllllllll@{}}
\toprule
\multirow{2}{*}{Case} & \multirow{2}{*}{\#Stories} & \multirow{2}{1cm}{w/o reason} & \multicolumn{4}{c}{Size in Words} & \multicolumn{4}{c}{Findings absolute} & \multicolumn{4}{c}{Findings per 1,000 Words} \\
 &  &  & Total & Role & Feature & Reason & T. & Ro. & F. & Re.  & T. & Ro. & F. & Re. \\ \midrule
C1 & 168 & 23 & 5226 & 801 & 2375 & 2050 & 229 & 1 & 83 & 145 & 44 & 1 & 35 & 71 \\
C2 & 123 & 45 & 2742 & 260 & 1552 & 930 & 120 & 0 & 62 & 58 & 44 & 0 & 40 & 62 \\
C3 & 230 & 19 & 6863 & 824 & 3090 & 2949 & 233 & 5 & 81 & 147 & 34 & 6 & 26 & 50 \\
C4 & 561 & 203 & 13124 & 1188 & 8223 & 3713 & 572 & 0 & 307 & 265 & 44 & 0 & 37 & 71 \\\midrule
Sum & 1082 & 290 & 27955 & 3073 & \textbf{15240} & \textbf{9642} & 1154 & 6 & \textbf{533} & \textbf{615} & 41 & 2 & \textbf{35} & \textbf{64} \\ \bottomrule
\end{tabular}
\end{sidewaystable*}

Table~\ref{tbl:findings_in_parts}~shows the results of this analysis. The number of words is roughly distributed as follows: 11\% of the words of a user story describe the role, 55\% of the words describe the feature and 34\% describe the reason. Of the 1,082 user stories, 290 had no reason part at all. Due to this uneven distribution, similar as in the previous analyses, we  normalize the number of findings by the number of words in each part resulting in the \emph{number of findings per 1,000 words}.

%Table~\ref{tbl:findings_in_parts} displays the number of findings in each part: 
Only 1\% of the findings are located in the role part. In fact, when we inspected these findings, they were false positives due to the grammatical problems described in the previous section. The absence of findings in this section is expected, as this part of the user story only names the role and does not offer many chances for smells as described in Sect.~\ref{sec:smells_29148}. For the remainder, 46\% of the findings are located in the feature and 53\% are located in the reason part. In relation to its size, the difference is striking: With 64 findings per 1,000 words, the reason has nearly double the number of findings of the feature part and nearly 70\% more findings than the average requirement, as analyzed in Sect.~\ref{sec:rq1}.

%\paragraph{Answer to RQ~2.2.1}
In summary, the reason part of user stories is particularly prone to smells, but the qualitative analysis in RQ~2.2 reveals that practitioners consider findings in this section to be less relevant. This investigation could support further application of Requirements Smells in practice by helping to prioritize smells according to their location.

%treemaps einbauen ohne ENDE!

\subsubsection{RQ~3: Which requirements quality defects can be detected with smells?}
For 44 of the 51 requirements artifacts the students provided technical reviews. We qualitatively analyzed the results of 10 randomly selected reviews (around~20\%). The inspected reviews were conducted by 5--7 reviewers (mean:~5.6), took 90~minutes and resulted in 18--69~defects (mean:~38.1). We iterated through the 381 defects documented in the reviews and evaluated whether the smell detection produced findings indicating these defects. If no smell indicated the defect, we openly classified the defects. We did not quantify these results, because the resulting numbers would assume and suggest that the distribution of defects is representative for regular projects, which we are unsure about (i.e.\  because of a high number of spelling and grammatical issues).

The classification of the defects and their comparison with the detected smells resulted in the following list of of defects indicated by Requirements Smells: 

\begin{description}
\item [Sentence not understandable.] In some instances, when the defect suggested changing the sentence to improve understandability, these sentences were highlighted especially by the \texttt{vague pronouns} and \texttt{negative statements} smells.
\item [Improper legal binding.] Various requirements artifacts had issues with improper legal binding. In one case, the reviewers recognized this and demanded the use of the term \emph{must}. The \texttt{loopholes} smell pinpointed at this issue.
\item [Unspecified/unmeasurable NFRs.] Various smells, especially the \texttt{super\-latives smell}, indicated at defects of underspecification within non-functional requirements.
\end{description}

The remaining defects were not indicated by Requirements Smells. 

\paragraph{Interpretation}
The quantitative distribution of defects is not necessarily representative for industry projects and, thus, has not been not analyzed. The reviews clearly show that manual inspection discovered the same defects as in the previous research question: Understandability, legally binding terminology and underspecified requirements. These are issues with regards to representation but also the content described in the artifact. We argue that these issues are common for requirements artifacts. Requirements Smells can therefore indicate relevant defects from multiple, independent sources (manual inspection, interviews with practitioners, independent manual reviews) for multiple, independent cases. 

\paragraph{Answer to RQ~3}
Automatic smell detection can point to issues in both representation (e.g. improper legal binding) and content (underspecified/unmeasurable NFRs). The analysis of the reported defects indicates that more defects could be automatically detected (see section \emph{further discussion on detectability of defects} described next). Nevertheless, just as for static code analysis, we see that automatic analysis can not indicate all defects and thus must be accompanied by reviews~\cite{wagner2005comparing}. The fourth research question aims at analyzing this aspect in depth.

\begin{sidewaysfigure*}[htbp]
\begin{center}
\includegraphics[width=1\columnwidth]{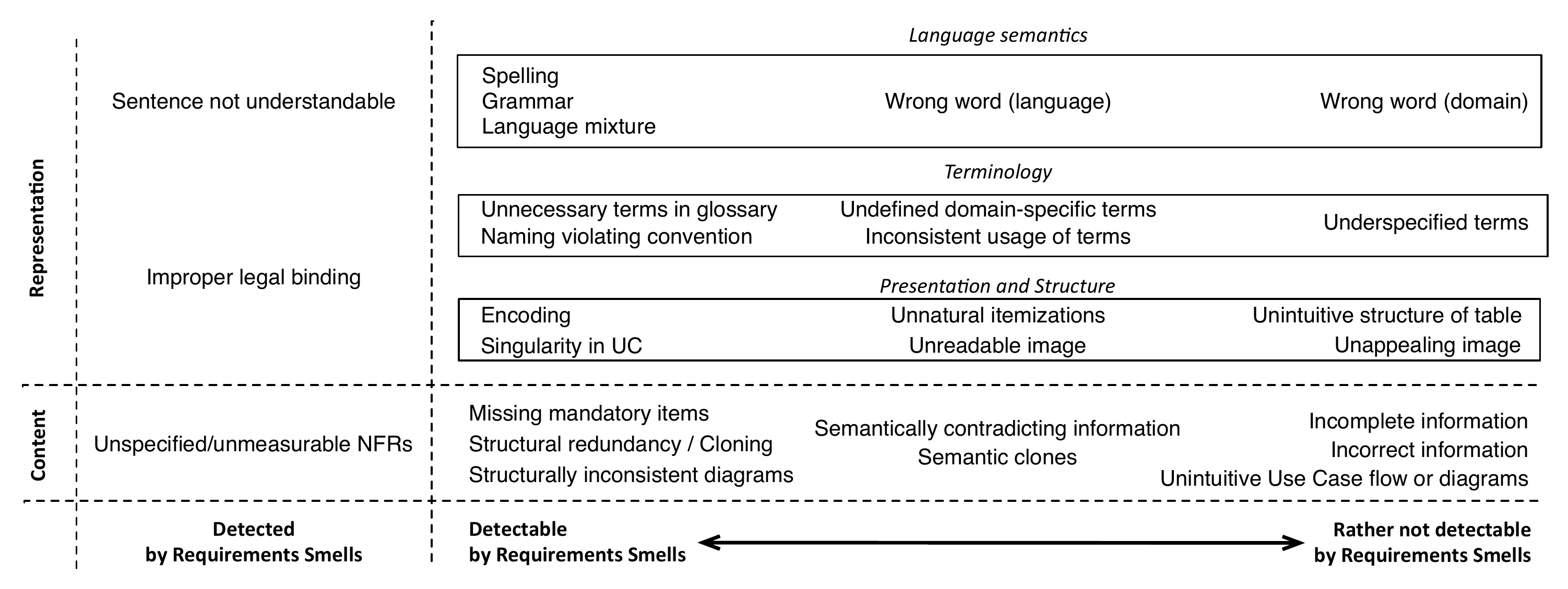}
\caption{Findings in requirements reviews, classified by content/representation and detection}
\label{fig:rq3_answer}
\end{center}
\end{sidewaysfigure*}

\paragraph{Further discussion on detectability of defects}
During the analysis, if no smells indicated the defect, we openly classified the defects. While discussing the resulting list of defects and the degree to which they are detectable within the group of authors, we came up with a classification which is broader as initially planned while designing the study. This classification considers whether a defect:  
\begin{compactitem}
\item \emph{Already can} be detected 
\item \emph{Could} be detected, but is not implemented yet in our detection 
\item \emph{Cannot be detected at the moment}, but should be soon
\item \emph{Cannot be detected at all} and probably won't be soon
\end{compactitem}
This classification is purely based on our knowledge of existing related work and our subjective expectations gained during the data analysis process. The classification yielded in a map visualised in Fig.~\ref{fig:rq3_answer}. The figure is structured in two dimensions: On the vertical axis, we group the defects into \emph{defects relating to the content}, and \emph{defects relating to representation}. Furthermore, on the horizontal axis, we map the items according to the expected precision and completeness we believe the detection could be (i.e. the classification above). The further left an item, the more precise and complete we expect a smell detection to be; the items on the right we assume to be close to impossible to detect in a general case. 

%We came to the conclusion that the defects can be grouped into the categories of language semantics, terminology defects, and presentation and structural defects. 

With the defects that our current approach does not reveal, this research question shows that more defects could be detected: These are namely defects with terminology, singularity in use cases and structural issues focusing on the content such as the absence of mandatory elements in the artifact~\cite{Kamata2007}, structural redundancy \cite{Juergens2010} or structural inconsistency between content. It remains unclear how far more enhanced language analysis with more sophisticated NLP and ontologies can enable to understand language. In any case, when a defect remains subtle and vague in its definition, such as an unintuitive structuring or design, we only see potential for automation if a defect can be defined precisely. For problems relating to the domain itself (e.g.\ incomplete information about the domain or incorrect information with regards to the domain), we consider it impossible to detect issues unless formalizing the concepts of the domain.

%Therefore, in our interpretation, the automatic detectability of the defects increases the more precisely the defect is defined on a syntactic level. 

\subsubsection{RQ~4: How could smells help in the QA process?}
%\todo{Dauer der analysen: war gering, deshalb gefragt ob anwendbar}

After the interviews and analysis, we asked all involved practitioners whether or not they think requirements smell detection is a helpful support, and whether and how they would integrate it in their context. We asked those questions openly and transcribed the answers for validation by the interviewees and later coding. In the following, we report on the results structured by topics. Where applicable, we provide the verbatim answers in relation to their cases (A, B or C). 

\paragraph{Overall Evaluation} In general, all practitioners agreed on the usefulness of the smell detection even if considering different perspectives that arise from their process setting. One practitioner (Case C) reports that he expects one benefit in using smell detection is that it would lead to a reduction of the time spent for effort estimations (in context of agile methods), as the product owner could benefit from the smell detection on the fly and, thus, avoid misinterpretations later.

%\begin{table}[h!]
%\caption{Quotes on Overall Evaluation}

\newcommand{\widthOfQuotes}{.8\columnwidth}

\begin{center}
\begin{tabular}{|c p{\widthOfQuotes}|}
\hline
\rowcolor{gray!50}
\multicolumn{2}{|c|}{\textbf{Quotes on Overall Evaluation}}\\
\hline
A. & \emph{``I think that smells can help to analyze a specification.''} \\
B. & \emph{``The method of Requirements Smells is a valuable extension in the area of requirements engineering and gives helpful input concerning the quality of specified requirements in early development phases.''}\\
C. & \emph{``I think such a smell detection is of high value to make sure that our team is confronted with already quality assured [user] stories. This can reduce the time in our effort estimations, because the product owner would directly notice on the fly what could lead to misinterpretations later.''}\\
\hline
\end{tabular}
\end{center}
%\label{tbl:quotesEvaluation}
%\end{table}
\paragraph{Integration into Process} When asked for how the practitioners would integrate the smell detection into their process setting, we got varying answers depending on the process. The practitioner relying more on rich process models (Case B) could imagine using a smell detection either as a support for the person writing the requirements or as part of a more fundamental QA method for the company. But also the practitioner relying more on the agile methods (Case C) could imagine using Requirements Smells as a support for the person writing the requirements or in context of analytical QA. In addition, one potential use is seen in context of problem management. Importantly, all practitioners see the full potential of a smell detection only if integrated in their existing tool chain (see also quotes on constraints and limitations). 

\begin{center}
\begin{tabular}{|c p{\widthOfQuotes}|}
\rowcolor{gray!50}\hline
\multicolumn{2}{|c|}{\textbf{Quotes on Integration into Process}}\\
\hline
B. & \emph{``I like to compare Requirements Smells to the ``check spelling aid" known e.g. from Microsoft Word. So for me Requirements Smells are intuitive and lightweight and should be used and integrated within requirements engineering and quality assurance processes.''}\\
C. & \emph{``As a product owner, I would use a smell detection on the fly [...].
In addition, smell detection could help in analytical QA, as it could reveal when a problem occurs repeatedly, either in a project or in the company as a whole.''}
\\\hline
\end{tabular}
\end{center}

\paragraph{Constraints and Limitations} One facet we consider especially interesting when using qualitative data is the chance to reveal further fields of improvement. We therefore concentrate now on the constraints that would hamper the usage of a smell detection. One facet we believe to be important is that practitioners want to avoid additional effort when using smell detection in their context. Furthermore, the practitioner of Case A believes that the automatic smell detection requires a common understanding on the notion of RE quality. He further indicates that the smell detection should explicitly take into account that some criteria cannot be met at every stage of a project.

\begin{center}
\begin{tabular}{|c p{\widthOfQuotes}|}
\rowcolor{gray!50}\hline
\multicolumn{2}{|c|}{\textbf{Quotes on Constraints and Limitations}}\\
\hline
A. & \emph{``First, the people who need to write the specification received training which gives the required performance criteria. Second, abstraction levels must be taken into account during the smell detection process, since at higher abstraction levels different criteria cannot be met (e.g.\ vague pronouns or subjective language).''}\\
B. & \emph{``As a product owner, I would use a smell detection on the fly provided that it would not mean additional effort [such as by having to use another tool].''}
\\\hline
\end{tabular}
\end{center}

\paragraph{Answer to RQ~4}
Our practitioners provided a general agreement on potential benefits of using smell detection a quality assurance context. When asked how they would integrate the requirements smell detection, they see possibility for both analytical and constructive QA, provided, however, this integration would not increase the required effort, e.g. by integrating the detection into existing tool chains. 
%The investigation of further barriers is in scope of future work. 

%Even though this is just anecdotal and, thus, subjective evidence, it forms a first external impression which encourages us to invest more effort into the development of Requirements Smells and analyze the approach in more depth.

%
%\subsubsection{Discussion}
%\todo[inline]{If needed}
%\todo[inline]{If makes sense, integrate old stuff}
%\begin{itemize}
%	\item ISO~29148 Language Criteria as Smells
%	\item Different Processes? Agile etc?
%	\item Subject of Analysis? Which part to analyze etc?
%\end{itemize}

\subsubsection{Evaluation of validity}
\label{sec:threats}
%\todo[inline]{TODO Stefan}

We use the structure of threats to validity from \cite{runeson12} to discuss the evaluation of the validity of our study.

\paragraph{Construct validity}
In our evaluation, we analyzed Requirements Smells in the terms of false positives, relevance and relation to quality defects. There are threats that the understanding of these terms varies and, thus, the results are not repeatable. Yet, we are confident that our validity procedures described in Sect.~\ref{sec:ValidityProcedure} reduced this threat. For the false positives, we classified a subset of the findings independently, and afterwards compared (inter-rater agreement Cohen's kappa: 0.53) and discussed the results. We subsequently reclassified a different subset of findings again, which lead to an inter-rater agreement (Cohen's kappa) of 0.72. For the classification of false negatives, we reclassified one document separately, calculating the percentage of agreement on false positives\footnote{We did not employ Cohen's kappa here, since the number of true positives (non-smell words) would strongly dominate the result and therefore skew the inter-rater agreement. Instead, we calculated the ratio of findings which both rating teams independently classified as false positive to the number of findings which only one of the teams classified false positive.}. This lead to an agreement of 88\%. 

We consider both of these substantial agreements, especially in the inherently ambiguous and complex domain of RE. Thus, we consider this threat as sufficiently controlled. 

%A further threat to the construct validity for the results of RQ~1 is that we might have missed further findings for smells. We cannot calculate a recall, however, because we would need to check the whole corpus of documents manually. This is beyond our capacity, and it would be likely that we miss findings as well. A higher number of findings for smells would only emphasize our conclusions. We also discarded a calculation of recall on a sample of the documents, because we observed difficulties in the differentiation between some of the Smells (in particular on subjective language vs.\ ambiguous adjectives and adverbs). Therefore, we need first an improved definition for some of the Smells which is in scope of future work.
%\todo[inline]{@Stefan/@Daniel: Recall discussion, maybe also in other sections of the paper}
%\todo[inline]{(HF:) Wheres for each finding, the inter rater agreement shows that it is possible to decide whether it is a valid finding, we found it harder to determine}

\paragraph{Internal validity}
A threat to the internal validity of our results is that the experience of the students as well as the practitioners might
play a role in their ratings of relevance or detection of quality defects. We mitigated this threat by choosing only practitioners
for the ratings and interviews who had several years of experience. The students are only in the second year. We cannot
mitigate this threat but consider the effect to be small. There might be some defects not found by the students that could have been 
indicated by a smell as well as unfound defects undetectable by smells. Hence, future studies will add to the classification
but are unlikely to change it substantially. Personal pride could potentially have an impact on the answers to a RQ~2.2, if practitioners are not able to professionally discuss their own work products. In our cases, however, all practitioners openly accepted the discussions (as can be seen in their answers). Even though we carefully supervised this threat, we have not found signs of personal bias in the cases involved. Finally, the students might also have been influenced by the review guidelines we provided. Yet, none of the investigated smells was explicitly listed 
in the guidelines. Instead, the guideline contained rather high-level aspects such as ``unambiguity''. Although we consider this threat to be a minor one, it is still present.

\paragraph{External validity}
As requirements engineering is a diverse field, the main threat to the external validity of our results is that we do not
cover all domains and ways of specifying requirements. We mitigated this threat to some degree by covering at least several
different domains and study objects, of which some are purely textual requirements artifacts, some use cases, and some user stories. We argue that this represents
a large share of today's requirements practices.

\paragraph{Reliability}
Our study contains several classifications and ratings performed by people. This constitutes a threat to the reliability of
our results. We are confident, however, that the peer debriefing and member checking procedures helped to reduce this
threat.

%%%%%%%%%%%%%%%%%%%%%%%%%%%%%%%%%%%%%%%%%%%%%%%%%%%%%%%%%%%%%%%%%%%%%%%

\section{Conclusion}
\label{sec:conclusion}

In this paper, we defined Requirements Smells and presented an approach to the detection of Requirements Smells which we empirically evaluated in a multi-case study. In the following,
we summarize our conclusions, relate it to existing evidence on the detection of natural language quality defects in requirements artifacts, and we discuss the impact and
limitations of our approach and its evaluation. We close with outlining future work.

\subsection{Summary of conclusions}
First, we proposed a light-weight approach to detect Requirements Smells. It is based on the natural language criteria of ISO~29148 and serves to rapidly 
detect Requirements Smells. We define the term \emph{Requirement Smell} as an indicator of a quality violation, which may lead to a defect, with a concrete location and a detection mechanism, and we also give definitions of a concrete set of smells. 

Second, we developed an implementation that is able to detect Requirements Smells by using part-of-speech (POS) tagging, morphological analysis and dictionaries. We found that it is possible to provide such tool support and outlined how such a tool could be integrated into quality assurance. 

Third, in the empirical evaluation, our approach showed to support us in automatically analysing requirements of the size of 250k words. Findings were present throughout all cases but in varying frequencies between 22 and 67 findings per 1,000 words. Outliers indicated serious issues. An investigation of the detection precision showed an average precision around 0.59 over all smells, again varying between 0.26 and 0.96. The recall was on average 0.82, but also varied between 0.5 and 0.95. To improve the accuracy, we described concrete improvement potential based on real world, practical examples. 

A further analysis of reviews and practitioner's opinions strengthen our confidence that smells indicate quality defects in requirements. For these quality defects, practitioners explicitly stated the negative impact of discovered findings on estimation and implementation in projects. The study also showed, however, that while Requirements Smell detection can help during QA presumedly in a broad spectrum of methodologies followed (including agile ones), the relevance of Requirements Smells varies between cases. Hence, it is necessary to tailor the detection to the context of a project or company. We analyzed this factor in depth, demonstrating that the reason part of a user story contains most findings (absolutely and relatively), but practitioners consider these findings less relevant as they argue that this part is not commonly used in implementation or testing. This raises the question of the relevance of this part at all, at least from a quality assurance perspective, which should be investigated in future work.

Our comparison with defects found in reviews furthermore showed that the Requirements Smell detection partly overlaps with results from reviews. As a result, we provide a map of defects in requirements artifacts in which we give a first indication where Requirements Smells can provide support and where they cannot.

Therefore, we provide empirical evidence from multiple, independent sources (manual inspection, interviews with practitioners, independent manual reviews) for multiple, independent cases, showing that Requirements Smells can indicate relevant defects across different forms of requirements, different domains, and different methodologies followed.

%\todo[inline]{@Henning: Discuss fuzziness, differentiation, etc. between smells?}

\subsection{Relation to existing evidence}
Existing approaches in the direction of automatic QA for RE are based on various quality models, including the ambiguity handbook by Berry et al.~\cite{Berry2003}, the now superseeded IEEE~830 standard~\cite{IEEEComputerSociety1998} and proprietary models. Yet, according to a recent literature review by Schneider and Berenbach~\cite{Schneider2013}, ISO~29148 is the current standard in RE \emph{``that every requirements engineer should be familiar with''}. However, no detailed empirical studies (see Table~\ref{tbl:identical_smells}) exist for the quality violations described in ISO~29148.
When comparing to similar, related quality violations, also few empirical, industrial case studies exist (see Table~\ref{tbl:related_work_eval}). Gleich et al.~\cite{Gleich2010} and Chantree et al.~\cite{Chantree2006} report for conceptually similar problems, a precision of the detection between 34\% and 75\% (97\% in a special case), and a recall between 2\% and 86\%. Krisch and Houdek~\cite{Krisch2015} report a lower precision in an industrial setting. 
The precision and recall for the detection of the smells, which we developed based on the description in the standard, are in a similar range to the aforementioned. 
In summary, this work provides a detailed empirical evaluation on the quality factors of ISO~29148, including a deeper understanding of both existing and novel factors.

We also take a first step from the opposite perspective: So far, to all our knowledge, all related work starts from a certain quality model and goes into automation. Our results to RQ~3 provides a bigger picture for understanding in how far quality defects in requirements could be addressed through automatic analysis in general.

Our results to RQ~2.2 furthermore provides evidence for the claim by Gervasi and Nuseibeh~\cite{Gervasi2002} that \emph{``Lightweight validation can discover subtle errors in requirements.''}  More precisely, our work indicates that automatic analysis can find a set of relevant defects in requirements artifacts by providing evidence from multiple case studies in various domains and approaches. The responses by practitioners to the findings do, to some extent, contradict the claim by Kiyavitskaya et al.~\cite{Kiyavitskaya2008} who state that ``\emph{any tool [...] should have 100\%~recall}''. Practitioners responded very positively on our first prototype and the smells it finds. Yet, obviously, more detailed and broader evaluations, especially conducted independently by other researchers not involved in the development of Smella, should follow.

\subsection{Impact/Implications}
For practitioners, Requirements Smells provide a way to find certain issues in a requirements artifact without expensive review cycles. We see three main benefits of this approach: First, the approach, just as static analysis for code, can enable project leads to keep a basic hygiene for their requirements artifacts. Second, the review team can avoid discussing obvious issues and focus on the important, difficult, domain-specific aspects in the review itself. Third, the requirements engineers receive a tool for immediate feedback, which can help them to increase their awareness for certain quality aspects and establish common guidelines for requirements artifacts.

Yet, the low precision for some of the smells might cause unnecessary work checking and rejecting findings from the automatic smell detection. Hence, at least for now, it is advisable to concentrate on the highly accurate smells.

For researchers, this work sharpens the term Requirements Smell by providing a definition and a taxonomy. By implementing and rating concrete smell findings, we also came to the conclusion, however, that not all of the requirements defects from ISO/IEC/\-IEEE~29148 can be clearly distinguished as Requirements Smells. In particular, the difference between \emph{Subjective Language}, \emph{Ambiguous Adverbs and Adjectives}, \emph{Non-verifiable Terms}, and \emph{Loopholes} was not always clear to us during our investigations (see RQ~2.1). Therefore, we, as a community, can take our smell taxonomy as a starting point, but we also need to critically reflect on some smells to further refine the taxonomy. 

Finally, empirical evidence in RE is, in general, difficult to obtain because many concepts depend on subjectivity~\cite{MMFV14}. One issue increasing the level of difficulty in evidence-based research in RE remains that most requirements specifications are written in natural language. Therefore, they do not lend themselves for automated analyses. Requirements Smell detection provides us with a means to quantify the extent of certain defects in a large sample of requirements artifacts while explicitly taking into account the sensitivity of findings to their context. Hence, this allows us to consider a whole new spectrum of questions worth studying in an empirical manner.

\subsection{Limitations}
We concentrated on a first set of concrete Requirements Smells based on our interpretation of the sometimes imprecise language criteria of ISO/IEC/\-IEEE 29148. There are more smells, also with different characteristics than the ones we proposed and analyzed. In addition, even though we diversified our study objects over domains, methods and different types of requirements, we cannot generalize our findings to all applicable contexts. We therefore consider the presented results only a first step towards the continuous application of Requirements Smells in software engineering projects.

\subsection{Future work}
Our work  focuses on Requirements Smells based on ISO/IEC/\-IEEE 29148. Future work needs to clarify and extend this taxonomy based on related work and experience in practice. This also includes the development of other Requirements Smell detection techniques to increase our understanding about which defects can be revealed by Requirements Smells and which defects cannot. %Also an investigation of the recall of the different Smells is needed after a sharpening of the Smell definitions. %We cordially invite researchers and practitioners to participate in the endeavor.

Second, this first study gained first insights into the usefulness of Requirements Smells for QA. We furthermore sketched an integration of Requirements Smells into a QA process. Yet, a full integration and the consequences must be analyzed in depth. In particular, we need to understand whether smell detection as a supporting tool, similar to spell checking, as pointed out by on of our participants, enables requirements engineers to improve their requirements artifacts.

Lastly, Requirements Smells focus on the detection of issues in requirements artifacts. They require a thorough understanding of the impact of a quality defect, which is hence also part of the requirements smell taxonomy. This link must be carefully evaluated and analyzed in practice. Our preliminary works on this topic~\cite{Femmer2014,MFME15} provide first ideas in that direction.

%\todo[inline]{@Henning: Clarification of smells}
%\todo[inline]{Add the recall aspect?}
%%%%%%%%%%%%%%%%%%%%%%%%%%%%%%%%%%%%%%%%%%%%%%%%%%%%%%%%%%%%%%%%%%%%%%%

\section*{Acknowledgments}
We would like to thank Elmar Juergens, Michael Klose, Ilona Zimmer, Joerg Zimmer, Heike Frank, Jonas Eckhardt as well as the software engineering students of Stuttgart University for their support during the case studies and feedback on earlier drafts of this paper.

This work was performed within the project Q-Effekt; it was partially funded by the German Federal Ministry of Education and Research (BMBF) under grant no. 01IS15003 A-B. The authors assume responsibility for the content.
\section*{Bibliography}
\bibliographystyle{abbrv}
\bibliography{bib}

\appendix

\section{Requirements Checklist}
\begin{table*}[htbp]
\caption{Checklist for the students' requirements reviews. Created by Anke Drappa, Patricia Mandl-Striegnitz and Holger R\"oder based on \cite{cockburn2001} and \cite{ludewig2010}. Translated from German.}\label{tbl:checklist}%\scriptsize
\centering
\tiny
\begin{tabular}{p{13cm}}
\toprule
The document is well structured and easy to understand.\\
\midrule
All used terms are clearly defined and consistently used.\\
\midrule
All external interfaces are clearly defined.\\
\midrule
The level of detail is consistent throughout the document.\\
\midrule
The requirements are consistent and unambiguous.\\
\midrule
The defined requirements are consistent with the state of the art.\\
\midrule
All tasks and data have useful identifiers.\\
\midrule
Data is not defined redundantly.\\
\midrule
The defined relationships between data objects are necessary and sufficient.\\
\midrule
The specification of quality attributes is realistic, useful, quantifiable and unambiguous.\\
\midrule
The user interface is comfortable and easy to learn.\\
\midrule
The use case describes a behaviour of the system which is valuable and visible for the actor. \\
\midrule
The use case is described in a table which is  consistently used for the whole requirements specification.\\
\midrule
The use case has a unique ID.\\
\midrule
The use case has a unique and expressive name.\\
\midrule
The main actor's goal is described in an understandable way.\\
\midrule
All actors participating in the use case are specified.\\
\midrule
If there is more than one actor, the main actor is identified.\\
\midrule
The preconditions of the use case are specified.\\
\midrule
The postconditions for the use case are specified.\\
\midrule
It is clearly specified how the main actor triggers the main success scenario.\\
\midrule
The main success scenario has 3 to 9 steps.\\
\midrule
After the main success scenario, the postconditions hold.\\
\midrule
The main actor reaches their goal by the main success scenario.\\
\midrule
Each step is sequentially numbered.\\
\midrule
It is clear which actor is executing the step.\\
\midrule
The step does not describe details of the user interface.\\
\midrule
The step describes exactly one action of the acting actor.\\
\midrule
There are postconditions for each extension.\\
\midrule
It is clearly specified in which step the main success scenario deviates into an extension.\\
\midrule
The conditions for the deviation into an extension are clearly specified.\\
\midrule
After an extension, all postconditions for that extension hold.\\
\bottomrule
\end{tabular}
\end{table*}

\end{document}